\documentclass[12pt, twocolumn, trackchanges, twocolappendix]{aastex631}
\usepackage{color}
\usepackage{amsmath}
\usepackage{float}
\usepackage{gensymb}
\usepackage{longtable}
\usepackage{threeparttablex}
\usepackage{booktabs}
\usepackage{placeins}
\usepackage{comment}
\usepackage{url}

\accepted{\apj}

\begin{document}
\title{Constraining Near-Simultaneous Radio Emission from Short Gamma-ray Bursts using CHIME/FRB}
\shorttitle{}
\shortauthors{}

\author[0000-0002-8376-1563]{Alice P.~Curtin}
  \affiliation{Department of Physics, McGill University, 3600 rue University, Montr\'eal, QC H3A 2T8, Canada}
  \affiliation{Trottier Space Institute, McGill University, 3550 rue University, Montr\'eal, QC H3A 2A7, Canada}

\author[0009-0008-6718-172X]{Sloane ~Sirota}
  \affiliation{Department of Physics, McGill University, 3600 rue University, Montr\'eal, QC H3A 2T8, Canada}
  \affiliation{Trottier Space Institute, McGill University, 3550 rue University, Montr\'eal, QC H3A 2A7, Canada}

\author[0000-0001-9345-0307]{Victoria M.~Kaspi}
  \affiliation{Department of Physics, McGill University, 3600 rue University, Montr\'eal, QC H3A 2T8, Canada}
  \affiliation{Trottier Space Institute, McGill University, 3550 rue University, Montr\'eal, QC H3A 2A7, Canada}

\author[0000-0003-2548-2926]{Shriharsh P.~Tendulkar}
  \affiliation{Department of Astronomy and Astrophysics, Tata Institute of Fundamental Research, Mumbai, 400005, India}
  \affiliation{National Centre for Radio Astrophysics, Post Bag 3, Ganeshkhind, Pune, 411007, India}
  \affiliation{CIFAR Azrieli Global Scholars Program, MaRS Centre, West Tower, 661 University Ave, Suite 505, Toronto, ON, M5G 1M1 Canada}

\author[0000-0002-3615-3514]{Mohit ~Bhardwaj}
  \affiliation{McWilliams Center for Cosmology, Department of Physics, Carnegie Mellon University, Pittsburgh, PA 15213, USA}

  \author[0000-0001-6422-8125]{Amanda M.~Cook}
  \affiliation{David A.~Dunlap Department of Astronomy \& Astrophysics, University of Toronto, 50 St.~George Street, Toronto, ON M5S 3H4, Canada}
  \affiliation{Dunlap Institute for Astronomy \& Astrophysics, University of Toronto, 50 St.~George Street, Toronto, ON M5S 3H4, Canada}

  \author[0000-0002-7374-935X]{Wen-Fai ~Fong}
    \affiliation{Center for Interdisciplinary Exploration and Research in Astrophysics (CIERA) and Department of Physics and Astronomy, Northwestern University, Evanston, IL 60208, USA}

\author[0000-0002-3382-9558]{B.M. ~Gaensler}
  \affiliation{Department of Astronomy and Astrophysics, University of California Santa Cruz, Santa Cruz, CA 95064, USA}
  \affiliation{David A.~Dunlap Department of Astronomy \& Astrophysics, University of Toronto, 50 St.~George Street, Toronto, ON M5S 3H4, Canada}
  \affiliation{Dunlap Institute for Astronomy \& Astrophysics, University of Toronto, 50 St.~George Street, Toronto, ON M5S 3H4, Canada}

\author[0000-0002-7164-9507]{Robert A. ~Main}
    \affiliation{Department of Physics, McGill University, 3600 rue University, Montr\'eal, QC H3A 2T8, Canada}
  \affiliation{Trottier Space Institute, McGill University, 3550 rue University, Montr\'eal, QC H3A 2A7, Canada}

\author[0000-0002-4279-6946]{Kiyoshi W. ~Masui}
    \affiliation{MIT Kavli Institute for Astrophysics and Space Research, Massachusetts Institute of Technology, 77 Massachusetts Ave, Cambridge, MA 02139, USA}
    \affiliation{Department of Physics, Massachusetts Institute of Technology, 77 Massachusetts Ave, Cambridge, MA 02139, USA}

\author[0000-0002-2551-7554]{Daniele Michilli}
    \affiliation{MIT Kavli Institute for Astrophysics and Space Research, Massachusetts Institute of Technology, 77 Massachusetts Ave, Cambridge, MA 02139, USA}
    \affiliation{Department of Physics, Massachusetts Institute of Technology, 77 Massachusetts Ave, Cambridge, MA 02139, USA}

  \author[0000-0002-8897-1973]{Ayush ~Pandhi}
  \affiliation{David A.~Dunlap Department of Astronomy \& Astrophysics, University of Toronto, 50 St.~George Street, Toronto, ON M5S 3H4, Canada}
  \affiliation{Dunlap Institute for Astronomy \& Astrophysics, University of Toronto, 50 St.~George Street, Toronto, ON M5S 3H4, Canada}

\author[0000-0002-8912-0732]{Aaron B. ~Pearlman}
    \affiliation{Department of Physics, McGill University, 3600 rue University, Montr\'eal, QC H3A 2T8, Canada}
    \affiliation{Trottier Space Institute, McGill University, 3550 rue University, Montr\'eal, QC H3A 2A7, Canada}

\author[0000-0002-7374-7119]{Paul ~Scholz}
    \affiliation{Department of Physics and Astronomy, York University, 4700 Keele Street, Toronto, ON MJ3 1P3, Canada}
    \affiliation{Dunlap Institute for Astronomy \& Astrophysics, University of Toronto, 50 St.~George Street, Toronto, ON M5S 3H4, Canada}

\author[0000-0002-6823-2073]{Kaitlyn ~Shin}
    \affiliation{MIT Kavli Institute for Astrophysics and Space Research, Massachusetts Institute of Technology, 77 Massachusetts Ave, Cambridge, MA 02139, USA}
    \affiliation{Department of Physics, Massachusetts Institute of Technology, 77 Massachusetts Ave, Cambridge, MA 02139, USA}

\newcommand{\allacks}{
A.P.C is a Vanier Canada Graduate Scholar. M.B is a McWilliams fellow and an International Astronomical Union Gruber fellow. M.B. also receives support from the McWilliams seed grant and North American ALMA Science Center. A.M.C. is funded by an NSERC Doctoral Postgraduate Scholarship. W.F. gratefully acknowledges support by the David and Lucile Packard Foundation, the Alfred P. Sloan Foundation, and the Research Corporation for Science Advancement through Cottrell Scholar Award 28284. The Dunlap Institute is funded through an endowment established by the David Dunlap family and the University of Toronto. B.M.G. acknowledges the support of the Natural Sciences and Engineering Research Council of Canada (NSERC) through grant RGPIN-2022-03163, and of the Canada Research Chairs program. K.W.M. holds the Adam J. Burgasser Chair in Astrophysics and is supported by NSF grants (2008031, 2018490). A.P. is funded by the NSERC Canada Graduate Scholarships -- Doctoral program. A.B.P. is a Banting Fellow, a McGill Space Institute~(MSI) Fellow, and a Fonds de Recherche du Quebec -- Nature et Technologies~(FRQNT) postdoctoral fellow. S.P.T. is a CIFAR Azrieli Global Scholar in the Gravity and Extreme Universe Program. K.S. is supported by the NSF Graduate Research Fellowship Program.
}


\correspondingauthor{Alice P. Curtin}
\email{alice.curtin@mail.mcgill.ca}

\begin{abstract}
We use the Canadian Hydrogen Intensity Mapping Experiment (CHIME) Fast Radio Burst (FRB) Project to search for FRBs that are temporally and spatially coincident with gamma-ray bursts (GRBs) occurring between 2018 July 7 and 2023 August 3. We do not find any temporal (within 1 week) and spatial (within overlapping 3$\sigma$ localization regions) coincidences between any CHIME/FRB candidates and all GRBs with $1\sigma$ localization uncertainties $<1\degree$. As such, we use CHIME/FRB to constrain the possible FRB-like radio emission for 27 short gamma-ray bursts (SGRBs) that were within 17$\degree$ of CHIME/FRB's meridian at a point either 6 hrs prior up to 12 hrs after the high-energy emission. Two SGRBs, GRB 210909A and GRB 230208A, were above the horizon at CHIME at the time of their high-energy emission and we place some of the first constraints on simultaneous FRB-like radio emission from SGRBs. While neither of these two SGRBs have known redshifts, we construct a redshift range for each GRB based on their high-energy fluence and a derived SGRB energy distribution. For GRB 210909A, this redshift range corresponds to $ z = [0.009, 1.64]$ with a mean of $z=0.13$. Thus, for GRB 210909A, we constrain the radio luminosity at the time of the high-energy emission to $\textrm{L} <2 \times 10^{46}$ erg s$^{-1}$, $\textrm{L} <5 \times 10^{44}$ erg s$^{-1}$, and $\textrm{L} <3 \times 10^{42}$ erg s$^{-1}$ assuming redshifts of $z=0.85$, $z=0.16$, and $z=0.013$, respectively. We compare these constraints with the predicted simultaneous radio luminosities from different compact object merger models. 
\end{abstract}


\section{Introduction} 
\label{sec:intro}

Gamma-ray bursts (GRBs) are extremely energetic (isotropic equivalent luminosities $\sim 10^{47} - 10^{53}$ erg s$^{-1}$) extragalactic bursts of gamma-rays. Since their initial discovery in the 1960s \citep{1973GRBDetection}, over 8000 GRBs have been published, \citep[e.g., see][]{GRBweb}. GRBs are typically split into two classes: short (SGRBs) and long (LGRBs). Usually, the distinction between the two is based on the burst duration, e.g., SGRB's $T_{90}$\footnote{The $T_{90}$ of a GRB is defined as the duration of the burst containing 90\% of the bursts observed total energy.} is typically less than 2 s. A more thorough classification, however, requires additional spectral information. For example, SGRBs typically have harder spectra than LGRBs \citep[][]{1993Kouveliotou, 2009Ghirlanda, 2020Jespersen, 2023ApJ...945...67S}.

SGRBs are likely produced by compact object mergers such as black hole (BH)-neutron star (NS) mergers or NS-NS mergers \citep{2014Berger}. The link between NS-NS mergers and at least some SGRBs was solidified with the joint discovery of GW170817 and GRB\,170817A \citep{2017ApJ...848L..13A}.  However, some nearby (e.g., $\lesssim$ 10 Mpc) SGRBs have been observed from magnetar giant flares \citep[][]{pbg+05, 2021ApJ...907L..28B, 2021NatAs...5..385F, 2021Natur.589..211S}. LGRBs, on the other-hand, are likely produced during the collapse of a rapidly rotating massive star \citep{2019GalYam}. However, two LGRBs have been associated with a kilonova, suggesting that some LGRBs can be produced by compact object mergers \citep{2022LGRBKilonova, Yang2022, Troja2022, Gillanders2023, Levan2024}.

Multiple theories predict that GRBs may be accompanied by transient radio emission similar to a fast radio burst \citep[FRB; ][to name a few]{Wang2016, Zhang2020, Sridhar2021Wind, 2023MostPhilippov, Cooper2023}. FRBs are short timescale (nanoseconds - seconds), highly energetic (radio luminosities $\sim10^{36}-10^{44}$  erg s$^{-1}$) transient radio bursts originating from extragalactic distances. While the majority of FRBs are apparently non-repeating \citep[e.g., only detected once thus far;][]{chimefrbcatalog}, a small sub-sample have been seen to repeat \citep{ssh+16a,abb+19b,abb+19c,fab+20, RN3}, and two show periodic bursting activity \citep{aab+20, rms+20, Cruces2021}. 

The inferred beaming-corrected rate of SGRBs based on observations is at most 1800 Gpc$^{-3}$ yr$^{-1}$ \citep{Fong2015, escorial2022jet}. This is an order of magnitude beneath the rate inferred for FRBs of $7.3 \times 10^{4}$  Gpc$^{-3}$ yr$^{-1}$ assuming no beaming \citep{Shin2023}\footnote{The beaming-corrected FRB rate would be higher than this rate, and hence there would be an even larger gap between the FRB and SGRB rate.}. However, it is possible that there are different classes of FRBs (e.g., repeater vs. non-repeater), each with a different progenitor. Thus, a sub-sample of the FRB population could be associated with SGRBs. We note that while both FRBs and GRBs are highly energetic transients, FRB radio luminosities are $\sim3-17$ orders of magnitude lower than GRB gamma-ray luminosities. However, the radio efficiency $\epsilon$ is expected to be of order $10^{-1}$ or less based on observations of pulsar efficiencies \citep{1993ApJS...88..529T, HansenLyutikov2001, Wang2016}. Additionally, as the gamma-ray emission is produced after the merger while the radio would be produced pre-merger, the two may harness different energy sources.

For SGRBs produced in the merger of two compact objects, FRB-like bursts could be produced prior to merger at the fronts of accelerated winds \citep{UsovKatz2000, Sridhar2021Wind}, from interactions between NSs in a manner similar to a unipolar inductor \citep{2012Piro, Wang2016}, from electromagnetic flares interacting with an orbital current sheet \citep{2020MostPhilippov, 2022MostPhilippov, 2023MostPhilippov}, from magnetic braking similar to that of isolated pulsars \citep{PshirkovPostnov2010, Totani2013} or from other interactions between the two compact objects \citep{HansenLyutikov2001, Mingarelli2015, Zhang2016BHBH, Yamasaki2018, Wada2020, Zhang2020, Wada2020, Carrasco2021}. FRB-like emission is also possible after a compact object merger \citep{PshirkovPostnov2010, Totani2013, zhang2014, Mingarelli2015}, but it is unlikely that the surrounding medium after a merger would be transparent to radio emission until $\sim$years later \citep{Yamasaki2018, Bhardwaj2023}.

FRBs could also be associated with the sites of LGRBs produced during the collapse of a rapidly rotating massive star. However, there are currently few theoretical predictions for FRBs produced prior to the LGRB, and the resulting supernova remnant would likely be opaque to radio emission for $\sim$ decades after the LGRB \citep{mbm17}. If a magnetar remnant is produced, we could possibly detect radio emission from this central magnetar $\sim$decades after the LGRB \citep{mbm17}.

There have been multiple efforts to search for FRB-like radio emission associated with GRBs \citep{2012Bannister, 2013Staley, Obenberger2014, Palaniswamy2014, Kaplan2015, Madison2019, Men2019,  2020Hilmarsson, 2021Anderson, 2021Rowlinson, 2021Bruni,  Curtin2023}, along with efforts to search for radio emission from magnetar giant flares \citep{tkp16, 2023ATelCurtin}.  Among all of these studies, there is only one possible association of an SGRB with a radio flare. \citet{rowlinson2023coherent} found a a tentative radio burst 76.6 minutes after GRB 201006A. The origin of this burst was further explored by \citet{Sarin2024}. If the radio burst is real and associated with the SGRB, then \citet{Sarin2024} conclude that a BH likely formed after the merger with the emission produced through synchrotron maser emission or magnetic reconnection within a jet. However, we note that the the probability of chance association between the radio flare and the SGRB is 0.5\% (2.6$\sigma$), and hence it remains an uncertain association. 

There have also been numerous high-energy follow-up observations of extragalactic FRBs, along with searches through archival high-energy catalogs for accompanying high-energy emission from extragalactic FRBs \citep[e.g., ][Cook et al. \textit{in prep.}]{sbh+17, Cunningham2019,Martone2019,  scholz2020, Guidorzi2020, 2020SwiftGuano, Casentini2020, Anumarlapudi2020, Verrecchia2021, Principe2021arXiv, Mereghetti2021, 2021Sakamoto, Pearlman2023M81}. Similarly, while there have been claims of possible associations \citep{2012Bannister, 2016Delaunay}, there has been no definitive association between an extragalactic FRB and a high-energy counterpart. 

The absence of detections in the above described searches can be attributed to several factors. First, the high-energy limits for FRBs are often above theoretical predictions, primarily due to the limited sensitivity of current high-energy instruments \citep{2020Chen}. Consequently, follow-up efforts have been confined to relatively nearby FRBs e.g., \citet{Pearlman2023M81}. Second, if FRBs belong to different classes, certain classes might not produce high-energy emission. For example, theories involving interactions between axion stars and compact bodies do not predict any accompanying high-energy emission \citep[]{AxionStar2017}. However, there are few other theories that do not predict a counterpart. Third, most searches for FRB-like counterparts to GRBs have been conducted significantly after the GRB, at a period when the medium surrounding the GRB site is likely opaque to radio waves \citep{Yamasaki2018, Bhardwaj2023}. Lastly, the FRB and its high-energy counterpart could be beamed in different directions \citep{Sridhar2021Wind}.

A promising instrument for overcoming some of these challenges is the Canadian Hydrogen Intensity Mapping Experiment (CHIME) FRB Project. CHIME/FRB's wide field of view (FOV; $\sim 250$ deg.$^2$; see Section \ref{section: chime overview}) and transit nature make it an ideal instrument for finding FRB-like emission associated with GRBs, as it can detect an FRB-like counterpart days prior to or days after the GRB. In \citet{Curtin2023}, we used CHIME/FRB to search for search for FRB-like radio emission associated with 81 GRBs detected between 2018 July 17 and 2019 July 8 by the \textit{Neil Gehrels Swift Observatory} Burst Alert Telescope (\textit{Swift}/BAT) and the \textit{Fermi} Gamma-ray Burst Monitor (\textit{Fermi}/GBM). We did not find any temporally (within 1 one week) and spatially (within 3$\sigma$) FRB and GRB coincidences. As such, we developed an algorithm that uses CHIME/FRB to constrain FRB-like radio emission before, at the time of, and after a GRB.

Here, we extend the time period used in \citet{Curtin2023} by $\sim$ four years and use the CHIME/FRB experiment to search for coincident FRBs and GRBs between 2018 July 17 and 2023 August 3. We then use our previously developed algorithm to constrain the FRB-like radio emission from 27 SGRBs. We start in Section \ref{section: chime overview} with a brief overview of CHIME/FRB. In Section \ref{sec: sources}, we present our sample of FRBs and GRBs. In Section \ref{sec: SGRB search}, we discuss our search for accompanying FRB-like radio emission. In Section \ref{sec: upper limits section}, we use CHIME/FRB to constrain the possible FRB-like radio emission from a sample of 27 SGRBs. We discuss our constraints in the context of various models in Section \ref{sec: discussion}. Finally, we summarize our work and discuss future avenues in Section \ref{sec: summary}.

\section{Overview of CHIME/FRB} 
\label{section: chime overview}
The CHIME telescope and its FRB backend have been discussed in detail by \citet{abb+18} and \cite{2022chimeoverview}. As a brief overview, CHIME consists of four 100-m by 20-m cylindrical, parabolic reflectors oriented in the north-south (N-S) direction. CHIME has a N-S FOV of $\sim120\degree$ and an east-west (E-W) FOV of $\sim1.3$ to 2.5$\degree$ (dependent on frequency, with 1.3$\degree$ corresponding to 800-MHz) for a total FOV of $\sim250$ deg.$^2$ \citep{nvp+17}. Each reflector is equipped with 256 dual-polarization feeds operating between 400 and 800 MHz. 

The 2048 antenna signals are digitized and fed into an FX-style correlator \citep{Chikada1984, 2022chimeoverview}. There, 256 N-S beams are formed through a spatial fast Fourier transform of the antenna signals. Three more sets of 256 N-S beams are then formed for a total of 4 E-W rows, each consisting of 256 N-S beams. For more details on these formed beams, see \citet{nvp+17}, \citet{msn+19}, and \citet{Bridget2023FluxCalibration}. The beamformed data are then sent to the CHIME/FRB backend. Through a series of processing levels, radio frequency interference is removed \citep[see ][for more details on the radio frequency intereference removal process]{2022MasoudRFI}, the data are de-dispersed, and candidates with FRB-like signals are identified and recorded. The final set of FRBs published \citep[see ][for the first published CHIME/FRB catalog]{chimefrbcatalog} are verified by team members in near real time and then recorded in our databases. 

\section{Sources} \label{sec: sources}
\subsection{FRBs} \label{subsec: FRB sources}
Our FRB sample consists of 4306 CHIME/FRB candidates\footnote{We use the term candidates here as the majority of these events have not yet been published. While all candidates have been classified as an FRB by at least two members of the CHIME/FRB collaboration, some of these candidates, from experience, may in reality consist of pulsar pulses, RFI, or events which are too faint to conclusively publish as an FRB.} discovered between 2018 July 7 and 2023 August 3. Many of these FRBs have already been published by \citet{chimefrbcatalog} and \citet{RN3}, or have been released through the CHIME/FRB VOevent service\footnote{\url{https://www.chime-frb.ca/voevents}}. Recently, CHIME/FRB published the channelized voltage data (herein referred to as baseband data) for 140 of the 536 FRBs published in the first CHIME/FRB catalog \citep{chimefrbcatalog, 2023ChimeBasebandCatalog}. With baseband data, it is possible to apply phase delays such that the telescope's response is maximized in a certain direction, a process known as beamforming. Utilizing this process, we mapped the signal's intensity around the initial source positions measured by the real-time pipeline \citep{2021DanielleBaseband}. This technique greatly improved the localization precision, yielding localization uncertainties $<1'$ \citep{2023ChimeBasebandCatalog}. This is a significant improvement over the real-time detection pipeline localizations (herein referred to as header localizations) published by \citet{chimefrbcatalog}. Thus, when possible, we also include the published high-time resolution data for bursts from the first CHIME/FRB catalog \citep{2023ChimeBasebandCatalog}.

\subsection{SGRBs} \label{subsec: SGRB sources}
Our GRB samples are compiled using two databases: the online  GRBWeb database\footnote{\url{https://user-web.icecube.wisc.edu/~grbweb_public/};
Access date October 10, 2023} \citep{GRBweb} and the General Coordinates Network (GCN)\footnote{\url{gcn.nasa.gov}}. GRBWeb is an excellent  tool for our study as it combines information from \textit{Fermi}/GBM \citep[][]{Meegan2009}, \textit{Swift}/BAT \citep{Gehrels2004, Barthelmy2005}, the INTErnational Gamma-Ray Astrophysics Laboratory \citep[\textit{INTEGRAL};][]{Winkler2003}, Konus-\textit{Wind} \citep{Aptekar1995}, BeppoSax \citep{beppo}, and the Burst and Transient Source Experiment (BATSE) on the Compton Gamma-ray Observatory \citep{1992Natur.355..143M}. If a given GRB is detected by multiple instruments, GRBWeb includes the localization region with the smallest uncertainty in its final table. We highlight, however, that GRBWeb is incomplete over certain times, namely in the period from 2003-2007. Hence we supplement GRBWeb's catalog with the GCN notices of \textit{Fermi}/GBM, \textit{Swift}/BAT, \textit{INTEGRAL}, and Konus-\textit{Wind}.

We further refine this larger GRB sample using criteria such as the localization area, duration ($T_{90}$), and detection date of the GRB. Depending on the given analysis (e.g., searching for associated GRBs and FRBs versus constraining FRB-like radio emission), we employ slightly different criteria. In total, we construct four different GRB samples. We list these four samples in Table \ref{table:SGRB samples} and discuss them in detail below.

\subsubsection{GRB Sample for Temporal and Spatial Coincidence Search}
In Section \ref{subsec: Intensity search}, we search for temporally (arrival times within a week of each other) and spatially (3$\sigma$ localization regions overlapping) coincident FRBs and GRBs. For this sample, we employ a selection criteria of a 1$\sigma$ positional uncertainty of $<1\degree$ for our GRBs. We also limit our sample to GRBs detected after 2018 July 17, seven days prior to the start of the pre-commissioning period of CHIME/FRB. Ultimately, this sample consists of 468 GRBs detected between 2018 July 17 and 2023 August 3 with an average localization uncertainty of 0.06$\degree$. We include both SGRBs and LGRBs in this sample. We define this as `Sample 1'. 

There are two GRBs (GRB 210909A and GRB 230208A) which were above the horizon at CHIME/FRB at the time of their high-energy emission Motivated by the fact that these GRBs have localization regions $>1\degree$ but $<3\degree$, we also re-do our temporal and spatial search using GRBs for which the localization region is $<3\degree$. As the probability of having a coincident FRB and GRB solely due to chance increases with increasing spatial uncertainties and increasing temporal separation, we only perform this search for the period of six hrs prior to up 12 hrs after the time of the GRB. This time frame is also consistent the time frame over which we calculate the radio upper limits (see Section \ref{sec: upper limits section}). This sample, defined as `Sample 2', consists of 723 GRBs (both SGRBs and LGRBs) between 2018 July 17 and 2023 August 3. The average localization uncertainty for this sample of GRBs is 0.74$\degree$. 

\subsubsection{GRB Sample for Solely Spatial Coincidence Search}
In Section \ref{subsec: baseband search}, we search for solely spatial coincidences with the 140 FRBs that have baseband data as the localization uncertainties for these FRBs ($<1'$) are significantly smaller than the header localization regions ($\sim$degs.). Given the solely spatial nature of this search, we extend our sample to all GRBs detected between 1991 April 21 and 2023 August 3. We then limit this sample to those GRBs with localization regions $<1\degree$. This sample consists of 2482 GRBs with an average localization uncertainty of 0.14\degree. We define this sample as `Sample 3.'

\subsubsection{GRB Sample for Upper Limits}
In Section \ref{sec: upper limits section}, we constrain the FRB-like radio emission from 27 SGRBs. Starting with Sample 2, we limit it to only include SGRBs as there are no strong theoretical predictions for near-simultaneous transient radio emission associated with an LGRB.  We first distinguish between SGRBs and LGRBs using their $T_{90}$ information and then manually verify all SGRB classifications using the respective telescope catalogs. If no classification is given in these catalogs, we look at the spectral hardness presented in the GCN circulars to ensure proper classification. We find that GRB 210217A, while having a duration of 4.22 seconds as detected by the \textit{Swift} telescope, was detected by \textit{Fermi} with a duration of 1.024 seconds. Despite further work to determine the nature of this GRB, it still remains unknown  \citep{GRB210217A_class}. We choose to include it in our sample of SGRBs. This final sample, which we define as `Sample 4', consists of 269 SGRBs.


\begin{deluxetable*}{c c c l c l l}
\tablecaption{GRB Samples} \label{table:SGRB samples}
\startdata
\\
Sample & GRBs\tablenotemark{a} & Cutoff\tablenotemark{b} & Temporal & LGRBs\tablenotemark{d} & Usage & Section\tablenotemark{e} \\
 &  &  & Window\tablenotemark{c} &  &  &  \\
\hline
1 & 468 & $1\degree$ & \begin{tabular}[c]{@{}l@{}}2018 July 17 - \\  2023 August 3 \end{tabular} &Y & \begin{tabular}[c]{@{}l@{}}Temporal and spatial search \\  with CHIME/FRB Intensity data\tablenotemark{f} \end{tabular} & Sect. \ref{subsec: Intensity search}  \\
2 & 723 & $3\degree$ &  \begin{tabular}[c]{@{}l@{}}2018 July 17 - \\  2023 August 3 \end{tabular}&Y  & \begin{tabular}[c]{@{}l@{}}Temporal and spatial search \\  with CHIME/FRB Intensity data\tablenotemark{g}\end{tabular} & Sect. \ref{subsec: Intensity search} \\
3 & 2482 & $1\degree$ & \begin{tabular}[c]{@{}l@{}}1991 April 21- \\  2023 August 3 \end{tabular} &Y  & \begin{tabular}[c]{@{}l@{}}Solely spatial search \\  with CHIME/FRB Baseband data\end{tabular} & Sect. \ref{subsec: baseband search} \\
4 & 269 & $3\degree$ & \begin{tabular}[c]{@{}l@{}}2018 July 17 - \\  2023 August 3 \end{tabular} & N & Constraining FRB-like emission & Sect. \ref{sec: upper limits section} \\
\hline
\enddata
\tablenotetext{a}{Number of GRBs in the sample.}
\tablenotetext{b}{1$\sigma$ GRB Localization Uncertainty Cutoff.}
\tablenotetext{c}{Time frame over which the GRBs were detected.}
\tablenotetext{d}{Whether or not LGRBs are included in this sample.}
\tablenotetext{e}{Section in which this sample is used for analysis.}
\tablenotetext{f}{Temporal search is restricted to within 1 week prior up to 1 week after the GRB. }
\tablenotetext{g}{Temporal search is restricted to 6 hrs prior up to 12 hrs after the GRB.}
\end{deluxetable*}

\section{Searching for FRB-like Counterparts to GRBs} \label{sec: SGRB search}

\subsection{Temporal and Spatial Coincidence Search with FRB Intensity Data} \label{subsec: Intensity search}

We search the entire CHIME/FRB candidates database (4306 candidates) for any temporal and spatial coincidences with GRB Sample 1. This is an extension of the work presented by \citet{Curtin2023} in which we searched for coincidences between 536 FRBs \citep{chimefrbcatalog} and all known GRBs. For a GRB and FRB candidate to be considered coincident in this search, their arrival times must be separated by less than 1 week and their spatial localization regions must agree within 3$\sigma$. The timescale of one week is largely motivated by the fact that the false positive rate of overlapping 3$\sigma$ localization regions is $\gtrapprox5\%$ for a period of $>1$ week, making it challenging to associate two events \citep[e.g., see Fig. 1 in ][]{Curtin2023}. For our GRBs, we assume that the localization regions are Gaussian and that a 1$\sigma$ localization can be extrapolated to a 3$\sigma$ localization region\footnote{Note that assuming a 2D Gaussian, a 3$\sigma$ localization region corresponds to a 98.9\% confidence interval.}. While the \textit{Swift}/BAT point spread function (PSF) is approximately Gaussian (and hence the above assumption is reasonable), this is not the case for more complex localizations from the interplanetary network (IPN) or from \textit{Fermi}/GBM. This is a caveat of our work.

As discussed in \citet{chimefrbcatalog}, typical CHIME/FRB header localizations are highly complex regions and include the first side-lobe of CHIME/FRB's formed beams. Hence, as seen in Figure 6 of \citet{chimefrbcatalog}, a 3$\sigma$ header localization region can span up to $\sim 5 \degree$ in R.A. and $\sim 1\degree$ in Decl. while the reported uncertainties are $<1\degree$. Thus, for all of our FRBs, we conservatively assume a 1$\sigma$ localization uncertainty region of $2\degree$ in R.A. and $0.4\degree$ in Decl. (and hence a 3$\sigma$ uncertainty region of $6\degree$ by $1.2\degree$) to encompass the full header localization regions, including the sidelobes\footnote{The 1$\sigma$ uncertainties correspond to the radii in a 2D ellipse.}. 

Using an R.A. uncertainty of $2\degree$ and a Decl. uncertainty of $0.4\degree$ for our FRB sample, we find four FRB-GRB coincidences within a week. We manually check the true CHIME/FRB header localization regions for the FRBs coincident with these four GRBs and find that none of the GRBs are spatially coincident with the actual 3$\sigma$ header localization contours. Hence, we do not find any spatial and temporal (within one week) coincidences between the CHIME/FRB candidates sample and GRB Sample 1.

We also check if any of the Sample 1 SGRBs were detected in the far sidelobes of CHIME/FRB. A far sidelobe event is one that occurs at least several beam widths away from the meridian\footnote{The full width half max of the CHIME/FRB primary beam is $\sim$1.6$\degree$ at 400-MHz, and the full width tenth max is $\sim$3$\degree$ at 400-MHz.} \citep[see][for more details on the CHIME/FRB sidelobes]{HsiuHsien2023arxiv}. A far side-lobe FRB would likely have an incorrect initial localization and hence would have been missed in our above searches. Using models of CHIME/FRB's beams \citep{nvp+17, 2022chimeoverview}, we construct the possible side-lobe tracks for each GRB and check whether any CHIME/FRB candidates lie along these side-lobe arcs. For any coincidences, we confirm whether or not the candidate is a sidelobe detection by checking its spectrum as a sidelobe FRB has a distinct spectral signature \citep{HsiuHsien2023arxiv}. We search for coincident events within 12 hrs of the SGRBs and find one coincidence. However, the FRB does not show the spectral features of a side-lobe event and hence it is not a far side-lobe event.

We also re-run our search with GRB Sample 2, as many of these GRBs are used in our upper limits analysis (see Section \ref{sec: upper limits section}). Here, we limit the temporal range from 6 hrs prior through to 12 hrs after the high-energy emission as the increased localization uncertainty greatly increases the chance of a temporal and spatial coincidence at large times. The temporal range here matches the range over which we determine upper limits on FRB-like radio emission in Section \ref{sec: upper limits section}. We choose 6 hrs prior through to 12 hrs after for the upper limits to be consistent with the time frame chosen in \citet{Curtin2023}. \citet{Curtin2023} chose an aysymmetric time range as most pre-SGRB models for FRB-like bursts predict the emission in the $\sim$minutes to seconds pre-merger while post-merger emission could last indefinitely.

We find one coincidence between GRB 181119A (GCN trigger 564330742) and FRB 20181119D (a repeat burst from FRB 20121102A). GRB 181119A occurred 4 hrs and 53 minutes after FRB 20181119D and is classified as an LGRB with a temporal duration of 16.445 s and a localization uncertainty of $2.8\degree.$ 
Following the methods presented by \citet{Curtin2023}, we calculate the chance probability of a GRB occurring $\sim$ 5 hrs after an FRB. We simulate a set of 723 GRBs (the same number of GRBs as in Sample 2) distributed evenly on the sky. For each GRB, we randomly draw an uncertainty from the true distribution of GRB uncertainties for Sample 2. As a reference, the average localization uncertainty of Sample 2 is $0.74\degree$. We find there is a 46$\%$ chance of having a GRB occur within 5 hrs of one of our FRBs. Hence, this is not a statistically significant association.

Most models for SGRBs predict radio emission that precedes or is coincident with the high-energy emission. Hence, we are most interested in the SGRBs which are above the CHIME horizon at the time of their high-energy emission. In our sample of SGRBs, there are two SGRBs that are within the FOV of CHIME at the time of their high-energy emission: GRB 210909A and GRB 230308A. For these two SGRBs, we conduct a search for coincident subthreshold (S/N $<8$) events in the CHIME/FRB database within one week. In doing so, we find a subthreshold event that occurred 2 hrs before GRB 210909A and is within the 2$\sigma$ localization region of GRB 210909A. As this is a subthreshold trigger, the only information available is the initial localization and time of arrival. To find the likelihood of this coincidence being due to chance, we simulate 100 GRBs randomly positioned on the sky, requiring them to be above the CHIME/FRB horizon at the time of their high-energy emission. We choose to simulate only 100 GRBs due to the computational cost of cross-checking an event with the CHIME/FRB subthreshold event database. We set the uncertainty on each GRB's localization region to be that of GRB 210909A. We cross-check our simulated set of GRBs with the CHIME/FRB subthreshold event database and find that there is a 6\% chance of finding at least one sub-threshold event coincident with a GRB within 2 hrs. Hence, while the chance probability is low, it is not low enough to claim an association between the two events.

\subsection{Solely Spatial Coincidence Search with FRB Baseband Data} \label{subsec: baseband search}

We re-do our search for temporal and spatial coincidences using the updated baseband localizations of the FRBs published in first CHIME/FRB catalog \citep{chimefrbcatalog, 2023ChimeBasebandCatalog}. We do not find any spatial coincidences between FRB and GRB pairs occurring within one week of each other. This result is expected as \citet{Curtin2023} performed the same search for these FRBs using the spatially larger header localization regions published earlier in \citet{chimefrbcatalog}. 

We also search for solely spatial coincidences between this FRB sample and GRB Sample 3. We find 11 solely spatial coincidences between this full GRB sample and the sources presented by \citet{2023ChimeBasebandCatalog}. None of the 11 GRBs has a known redshift and so we cannot perform a comparison between the FRB's redshift (which can be estimated from its dispersion measure; discussed below) and the GRB's redshift. To determine if the spatial coincidences using the baseband dataset are significant, we perform a chance coincidence analysis similar to that presented by \citet{Curtin2023}. We simulate a sample of 2482 GRBs distributed evenly on the sky. For the positional uncertainties of the GRBs, we randomly draw an uncertainty from the true distribution of GRB uncertainties for this sample. We then cross-check this sample of simulated GRBs with the 140 baseband localizations. We run 1000 simulations and find that there is an 86\% chance of having 11 or more spatially coincident FRBs and GRBs. As a reference, the average localization uncertainty of GRB Sample 4 is $0.14\degree$ while the average localization uncertainty of the 11 spatially coincident GRBs is $0.73\degree$. Hence, even using subarcminute spatial uncertainties for FRBs, we cannot claim a purely spatial association between any GRBs and FRBs unless the GRBs were to have far better positional uncertainties.

\section{Constraining FRB-like Radio Emission from SGRBs} \label{sec: upper limits section}

As we do not find any significant coincidences between our FRB and GRB samples, we switch to constraining the possible 400- to 800-MHz FRB-like radio flux and fluence for SGRBs within the FOV of CHIME/FRB at a point either 6 hrs before up to 12 hrs after their high-energy emission. We focus solely on SGRBs (GRB Sample 4) and define the FOV of CHIME/FRB as within $\sim 17 \degree$ of CHIME's meridian as the primary beam of CHIME has been well modelled within this range \citep{2022chimeoverview}. 

\subsection{Calculating Upper Limits on Radio Flux/Fluence using CHIME/FRB} \label{subsec: CalcUL}
To calculate upper limits on the radio flux and fluence, we follow the method outlined by \citet{Curtin2023}. As an overview, we use a previously detected FRB nearby in Decl. ($< 1 \degree$) to the SGRB as a flux calibrator. We do not require our FRB calibrator to be nearby in time, and instead account for temporal differences using daily CHIME/FRB system sensitivity metrics. We conservatively assume that CHIME/FRB is sensitive to bursts with a S/N threshold of 10 and then scale the ratio of the calibrator FRB's peak flux/fluence to S/N (e.g., $\frac{\textrm{flux}}{\textrm{S/N}})$ to a theoretical S/N of 10. The peak flux and fluence of the FRB are determined using the methodology described in \citet{Bridget2023FluxCalibration}. Only FRBs published in \citet{chimefrbcatalog} are used as calibrators.

Due to slight spatial differences between the positions of the calibrator FRB and the GRB of interest, we scale this ratio by a data-driven model of CHIME's primary beam \citep{2022chimeoverview} and an analytical model of CHIME/FRB's formed beams \citep{nvp+17} at the two locations. We assume a frequency range of 400- 800-MHz for the GRB. Finally, for any GRBs or FRB calibrators within 10$\degree$ of the Galactic plane, we scale the ratio by the sky temperature at the two locations using the 2014 all-sky continuum map at 408 MHz \citep{2015Haslam}. We assume an observing frequency of 600-MHz for CHIME/FRB and a brightness temperature spectral index of $-2.5$ for the Galactic synchrotron emission \citep{2003Bennett}. 

Due to the dispersion of radio waves travelling through an ionized medium, we must also determine the delay of the radio emission relative to the high-energy emission as this affects the relative arrival time of the emission at CHIME/FRB. This dispersive delay, quantified using the dispersion measure (DM), is the total delay along the entire GRB's line of sight (LOS) and hence consists of contributions from the Milky Way disk, the Milky Way's halo, the intergalactic medium, and the host galaxy of the GRB.  We calculate each component, and hence the total DM, in the same manner as in \citet{Curtin2023}.


Combining all of the above factors, the upper limit on the radio flux in the 400-
800-MHz CHIME/FRB band is given by: 

\begin{multline}
    \textrm{Flux}_{\textrm{GRB}} = 10 \times  \frac{\textrm{Flux}_{\textrm{FRB}}}{\textrm{S/N}_{\textrm{FRB}}} \times  \frac{\textrm{B}_{\textrm{M}}}{\textrm{B}_{\textrm{GRB}}} \times 
    \frac{1}{\textrm{F}_{\textrm{GRB}}} \\
    \times 
    \frac{\Delta\textrm{S}_{\textrm{Sys,GRB}}}{\Delta\textrm{S}_{\textrm{Sys,FRB}}} \textbf{\textrm{Jy}}
\label{eq: fluxUpperLimit}
\end{multline}

\noindent where \textbf{$\textrm{Flux}_{\textrm{FRB}}$ is the flux of the FRB in Jy}, {$\textrm{B}$} is the primary beam response \textbf{(unitless)}, $\textrm{F}$ is the formed beam response \textbf{(unitless)}, $\Delta\textrm{S}_{\textrm{Sys}}$ is the system sensitivity \textbf{(unitless)}, and $\textrm{M}$ stands for the response along the meridian \citep[see Eq. 2 of][]{Curtin2023}. We also scale our fluence limits to a fiducial burst width (W) of 10 ms e.g.,  $\textrm{Fluence}_{\textrm{GRB, Final}} = \textrm{Fluence}_{\textrm{GRB, Eq.1}} \times \sqrt{\textrm{W}_{\textrm{FRB}}/10\textrm{ ms}}$.

The uncertainties on our flux (fluence) upper limits are determined using the flux (fluence) uncertainties of our calibrator FRBs, the uncertainties of the daily CHIME sensitivity metrics, the uncertainties on our primary beam model, and the GRB's localization uncertainty. To account for the uncertainty in the GRB's localization, we sample using a Monte Carlo (MC) method over the GRB's 3$\sigma$ localization region and re-calculate the CHIME primary beam and formed beam sensitivities at 100 different locations within the GRB's 3$\sigma$ localization region\footnote{We note that again this is an oversimplification, as we assume that the 3$\sigma$ localization region is $3\times$ the reported 1$\sigma$ localization uncertainty.}.

\subsection{Results} \label{subsec: UL results}

\subsubsection{Constraints on Radio Fluxes \& Fluences}
\begin{figure*}
  \centering
   \includegraphics[width=0.9\textwidth]{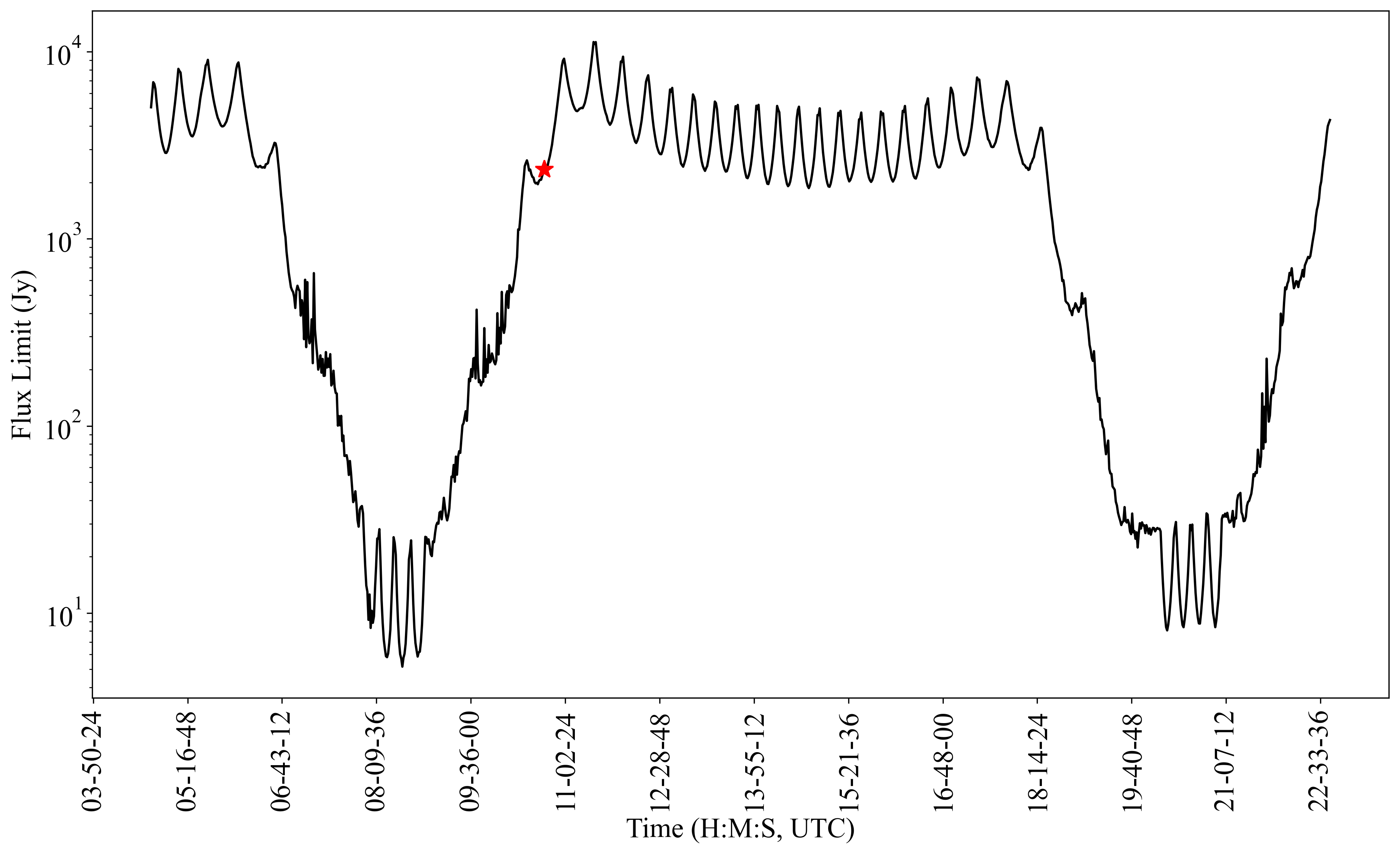}
  \caption{Radio flux upper limits on the 400- to 800-MHz radio flux for GRB 210909A assuming a 10-ms radio burst. Radio flux limits (99$\%$ confidence) before and after the high-energy emission (dispersion delay accounted for) are calculated every minute. The radio flux limit at the time of the high-energy emission (dispersion delay accounted for) is shown as a red star. The GRB was detected at 10:43:19 UT on 2021 September 9th (no dispersion correction) by the \textit{Fermi}/GBM instrument. The shape of the curve reflects the instrumental beam shape as GRB 210909A transits over CHIME in the band of 400- to 800-MHz, with the larger envelope due to changes in the primary beam sensitivity and the smaller ripples due to changes in the formed beam sensitivity at the position of the GRB. Smaller ripples are also due to the MC over the GRB's position (see Section \ref{subsec: CalcUL}). The four minima during each transit correspond to the transit over the four CHIME cylinders. Due to the high declination of this source, GRB 210909A transits over CHIME/FRB twice in a given day and hence the two dips in the flux limits shown here.}
  \label{fig:GRB210909A}
\end{figure*}

As discussed in Section \ref{subsec: SGRB sources}, there are two SGRBs that were above the CHIME horizon at the time of their high energy emission: GRB 210909A and GRB 230308A. GRB 210909A was detected by \textit{Fermi}/GBM, and with a Decl. of 83.85$\degree$, is within CHIME's FOV throughout the entire day \citep{2021GCN.30786....1F}. In Figure \ref{fig:GRB210909A}, we show the 3$\sigma$ flux upper limits for GRB 210909A as its position transits above CHIME. The red star indicates the FRB-like radio limits at the time of the high-energy emission while the black line traces the flux limits at other points in time. The shape of the curve does not reflect any inherent properties of the GRB, but instead reflects the instrumental beam shape as an object transits over CHIME in the band of 400- to 800-MHz, with the larger envelope due to changes in the primary beam sensitivity and the smaller ripples due to changes in the formed beam sensitivity at the position of the GRB. GRB 230308A is similarly at fairly high Decl of 66.61$\degree$ and is within CHIME's FOV\footnote{Defined here as within 17$\degree$ of meridian.} for 5.7 hrs. It was similarly detected by \textit{Fermi}/GBM \citep{2023GCN.33420....1F}. We present both SGRBs, along with our best constraints at the time of the high-energy emission, in Table \ref{table: Limits At Time GRB}. We note that a third GRB, GRB 201006A, was also above CHIME's horizon at the time of its high energy emission. However, CHIME was not operating with nominal sensitivity during this period and hence we do not include it in our work. 

\begin{deluxetable*}{c c c c c c c c c c c}
\tablecaption{CHIME/FRB Constraints on Coincident Radio Emission \label{table: Limits At Time GRB}}
\startdata
\\
Name
& HE Fluence\tablenotemark{a}
& Flux\tablenotemark{b}
& Fluence Ratio\tablenotemark{c}
& $\eta$\tablenotemark{d} 
& z$_{\textrm{low}}$ \tablenotemark{e}  
& z$_{\textrm{mean}}$ \tablenotemark{f}  
& z$_{\textrm{high}}$ \tablenotemark{g}  
& L$_{z_{\textrm{low}}}$\tablenotemark{h}  
& L$_{z_{\textrm{mean}}}$\tablenotemark{i} 
& L$_{z_{\textrm{high}}}$\tablenotemark{j}\\
& (erg$^{-1}$ cm$^2$) & (Jy) & ($10^8$ Jy ms & ($10^{-11}$) &  & & & (erg s$^{-1}$)& (erg s$^{-1}$) & (erg s$^{-1}$) \\
& & & erg$^{-1}$ cm$^2$) & & &  & & & &
\\ \hline
GRB 210909A & $3.0 \times 10^{-6}$ &$<$2000 & $<$1 & $<$50 & 0.013 & 0.16 & 0.85 &$<3 \times 10^{42}$& $<5 \times 10^{44}$ & $<2 \times 10^{46}$ \\ 
GRB 230308A & $7.7 \times 10^{-6}$ & $<$8000 & $<$130 & $<$5000 & 0.009 & 0.10 & 0.54 & $<6 \times 10^{42}$& $<8 \times 10^{44}$ & $<2 \times 10^{46}$ \\
\hline
\enddata
\tablenotetext{a}{Previously published high-energy fluence of the GRB.}

\tablenotetext{b}{Upper limit on the 400- to 800-MHz radio flux at the time of the high-energy emission (accounting for the estimated dispersion delay; see Section \ref{subsec: CalcUL}) at the 99\% confidence level assuming a 10-ms radio burst.}

\tablenotetext{c}{Same as $a$ except for the radio-to-high-energy fluence ratio.}

\tablenotetext{d}{Same as $a$ except for $\eta$ (dimensionless radio-to-high-energy fluence ratio assuming a 400-MHz radio bandwidth). The energy range for GRBs detected by \textit{Swift}/BAT is 15 to 150 keV and that for those detected by \textit{Fermi}/GBM is 10 to 1000 keV.}

\tablenotetext{e}{The minimum redshift assumed for this GRB based on the 90th percentile of the SGRB energy distribution.}

\tablenotetext{f}{The mean redshift assumed for this GRB based on the SGRB energy distribution.}

\tablenotetext{g}{The maximum redshift assumed for this GRB based on the 90th percentile of the SGRB energy distribution.}

\tablenotetext{h}{Same as $a$ except for the radio luminosity assuming a redshift of z$_{\textrm{low}}$.}

\tablenotetext{i}{Same as $d$ except for a redshift of z$_{\textrm{mean}}$.}

\tablenotetext{j}{Same as $d$ except for a redshift of z$_{\textrm{high}}$.}
\end{deluxetable*}

There are another 25 SGRBs that were above the horizon at a point either 6 hrs before up to 12 hrs after their high energy emission and for which the CHIME/FRB system was operating nominally during this period. Of the 27 SGRBs, 17 are \textit{Fermi}/GBM SGRBs and 10 are \textit{Swift}/BAT SGRBs. For each SGRB, we calculate upper limits starting 6 hrs prior to the high-energy emission through to 12 hrs after. However, for many of the SGRBs, we cannot calculate limits over the entire 18 hr period as most GRB positions are only within the FOV of CHIME for $\sim$minutes during a day. Only sources with a very high Decl.  e.g., GRB 210909A, remain within CHIME's FOV for multiple hrs. For each of the 27 SGRBs, we present our best flux and fluence ratio (radio to high-energy) constraints over the 18 hr period in Table \ref{table: individual GRBs}. 

\begin{figure*}
\gridline{\fig{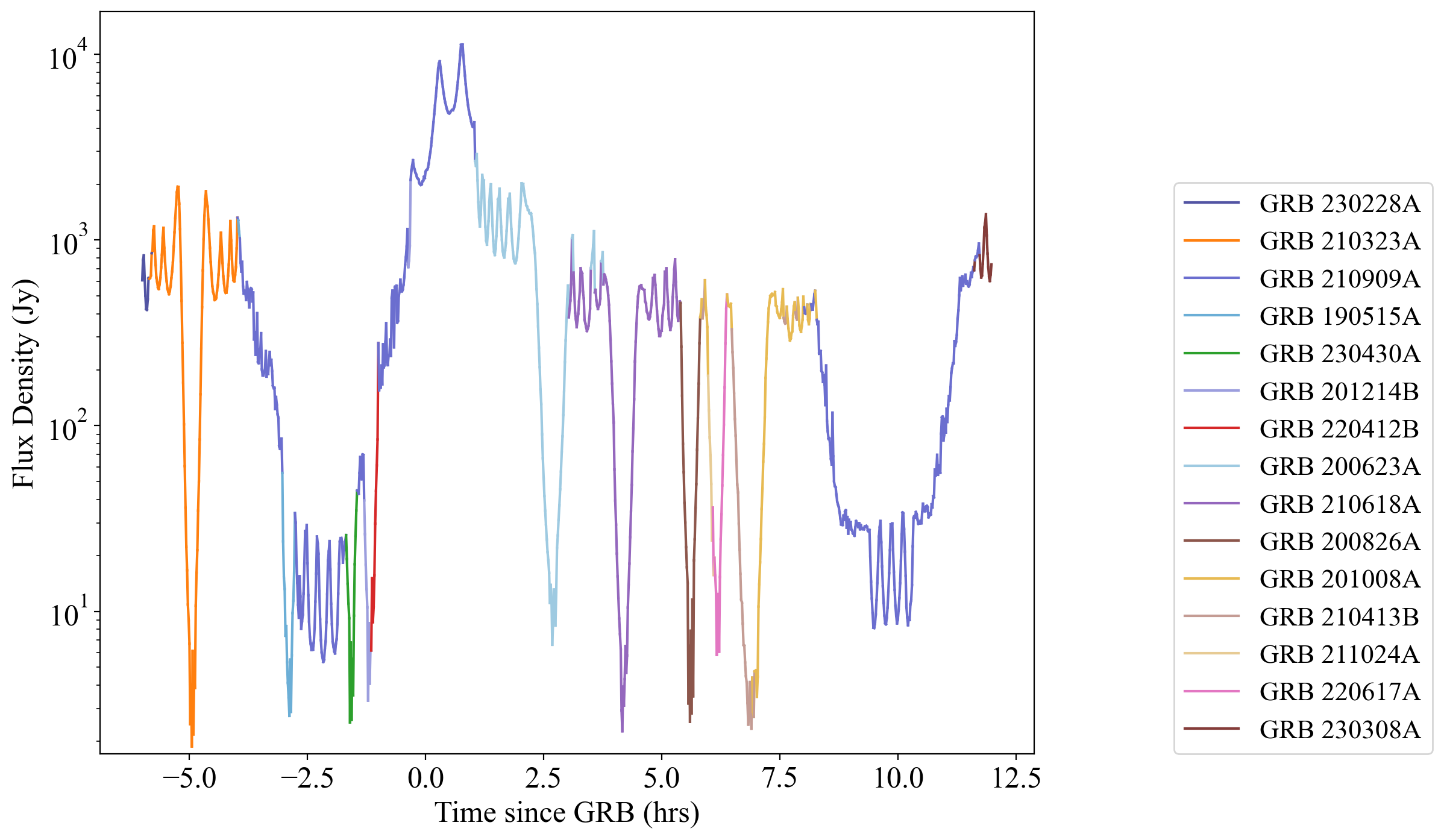}{0.98\textwidth}{}
          }
\gridline{\fig{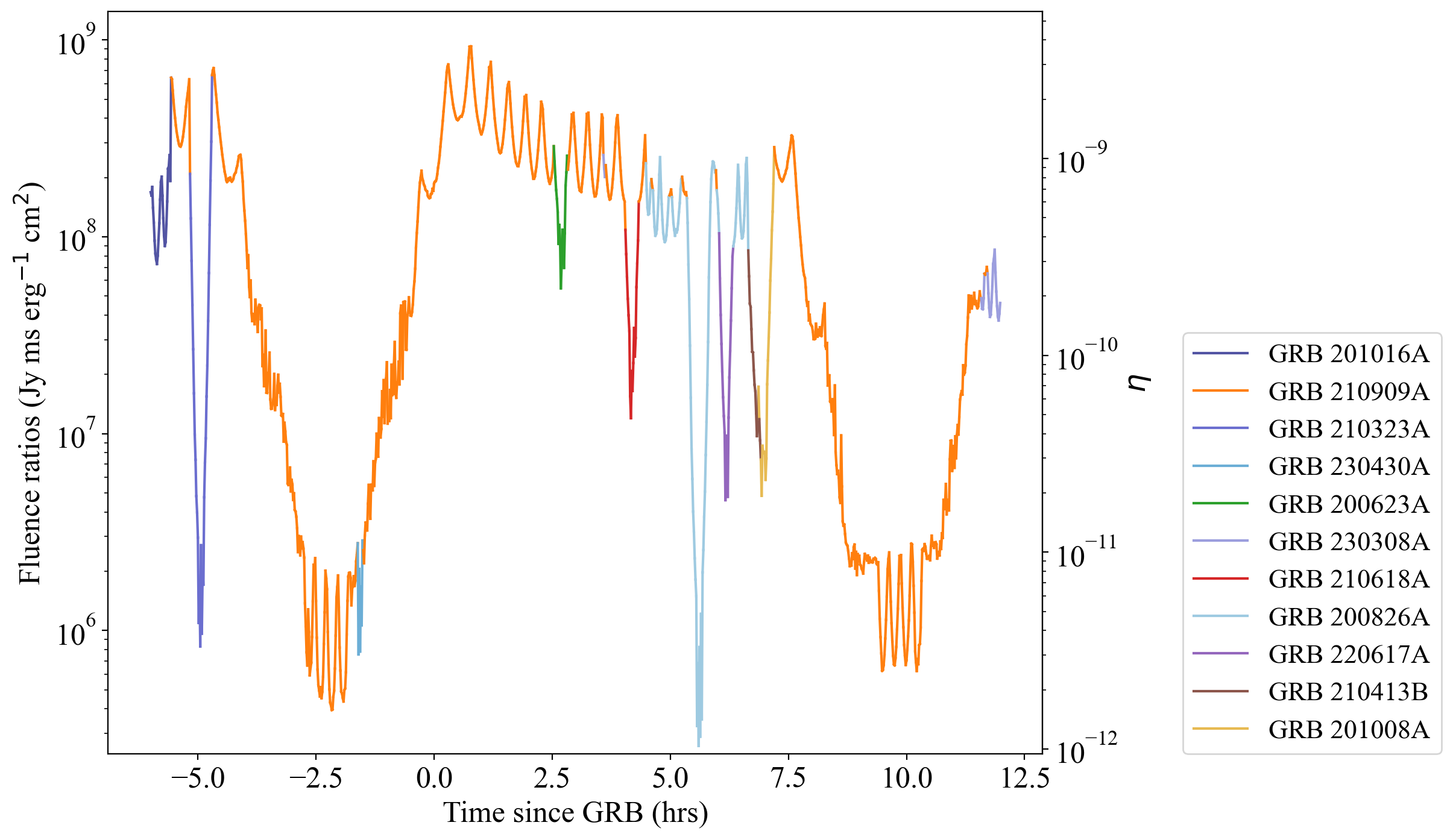}{0.98\textwidth}{}}
\caption{Radio flux (top) and fluence ratio (bottom) limits at the 99\% confidence level for SGRBs within our sample detected by \textit{Fermi}/GBM. Upper limits are calculated every minute starting six hrs prior up until 12 hrs after the SGRB, with the time shifted to account for the estimated DM delay. For each relative timestamp, we show our most constraining limit from the sample of SGRBs by \textit{Fermi}/GBM. The different SGRBs, along with the time periods over which they are used to constrain the phase space, are shown as different colors. The \textit{Fermi}/GBM energy range is 1 to 1000 keV. }
\label{fig:fermilimits}
\end{figure*}

\begin{figure*}
\gridline{\fig{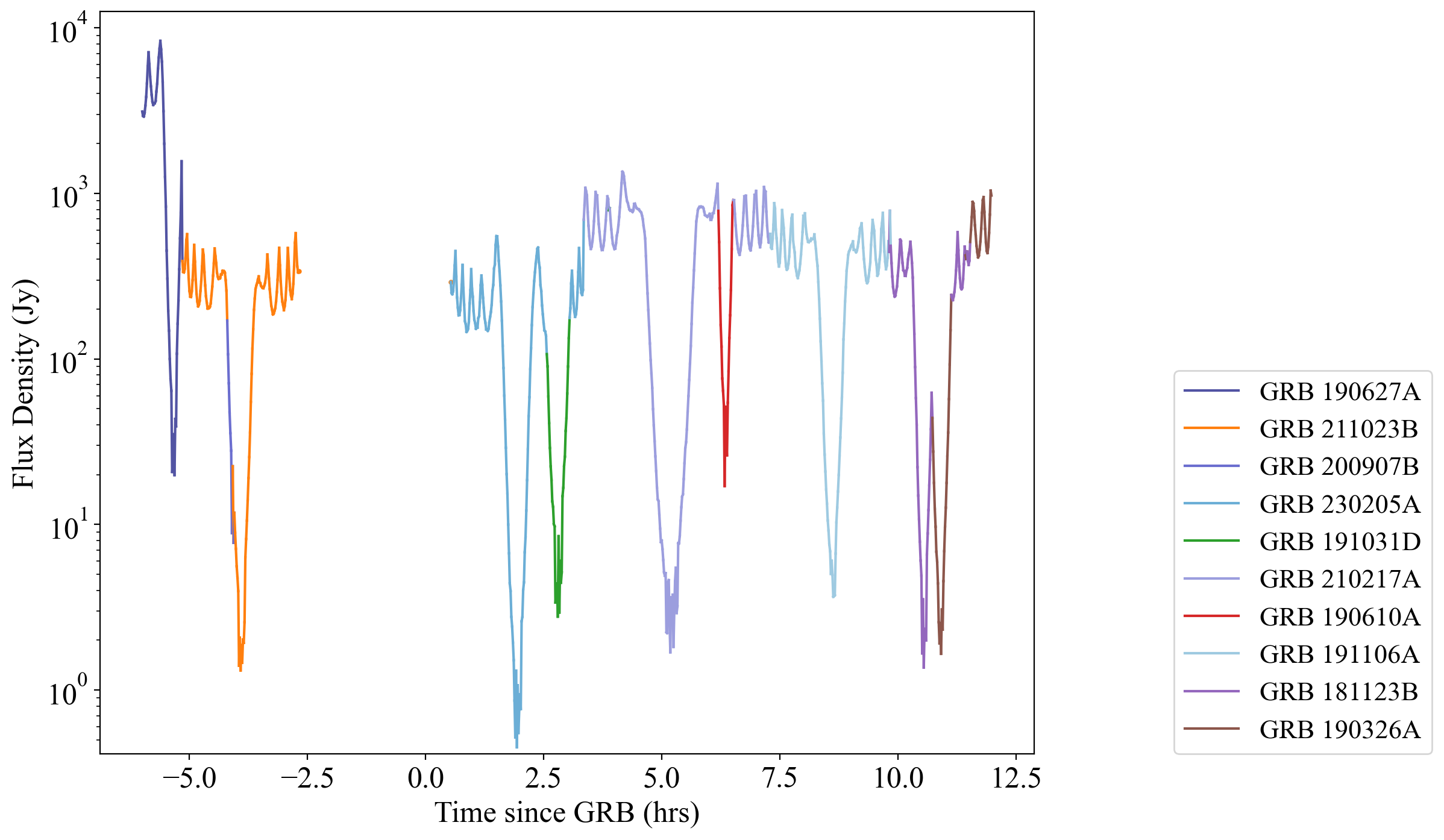}{0.98\textwidth}{}
          }
\gridline{\fig{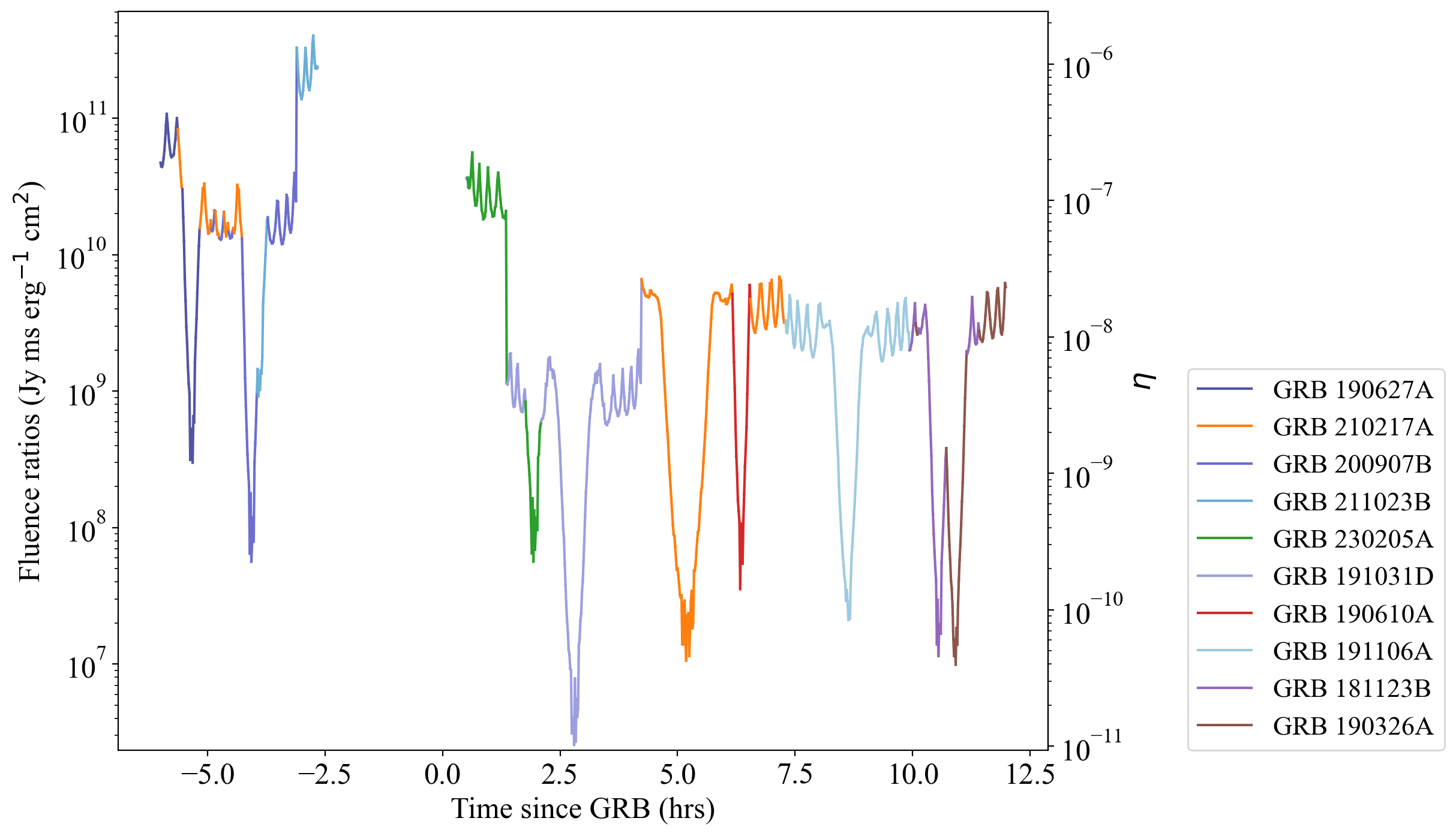}{0.98\textwidth}{}
        }
\caption{Same as Figure \ref{fig:fermilimits}, but for \textit{Swift}/BAT. The energy range for \textit{Swift}/BAT is 15 to 150 keV.}
\label{fig:swiftlimits}
\end{figure*}

If we assume that a possible radio counterpart to a GRB would be produced via the same mechanism for all of the SGRBs in our sample, then we can combine the limits from multiple SGRBs to create a larger picture of the constrained FRB-like radio emission. Thus, for each respective SGRB sample (e.g., \textit{Fermi}/GBM or \textit{Swift}/BAT SGRBs), we combine our radio limits (such as those shown in Figure \ref{fig:GRB210909A}) from multiple SGRBs to create a semi-continuous set of limits that begin 6 hrs prior to and extend to 12 hrs after the arrival time of the high-energy emission. We show our combined 3$\sigma$ flux and fluence ratio limits for SGRBs detected by \textit{Fermi}/GBM in Figure \ref{fig:fermilimits} as a function of time relative to the high-energy emission. We present the fluence ratio limits both in units of Jy ms erg$^{-1}$ cm$^2$ and as a dimensionless quantity $\eta$ assuming a 400-MHz bandwidth. The high-energy fluences are taken from the online \textit{Fermi}/GBM catalog \citep{Meegan2009}. The flux and fluence ratio limits for the SGRBs detected by \textit{Swift}/BAT are presented in Figure \ref{fig:swiftlimits}. The high-energy fluences for the SGRBs from \textit{Swift}/BAT are taken from the online \textit{Swift}/BAT catalog \citep{Gehrels2004, Barthelmy2005}. The gaps present in Figure \ref{fig:swiftlimits} are due to a lack of strong constraints for any SGRBs at these times.

\begin{table*}[p]
\begin{ThreePartTable}
\begin{TableNotes}
\item\tablenotetext{a}{Time (in hrs) before (negative time) or after the detected high-energy emission for which the radio flux/fluence limits apply. These times correct for an estimated dispersion delay (e.g., see Section \ref{subsec: CalcUL}).}

\tablenotetext{b}{Upper limit on the 400- to 800-MHz radio flux at the 99$\%$ confidence level for a 10-ms radio burst.}

\tablenotetext{c}{Same as $b$ except for the radio-to-high-energy fluence ratio.}

\tablenotetext{d}{Same as $b$ except for $\eta$ (dimensionless radio-to-high-energy fluence ratio assuming a 400-MHz radio bandwidth). The energy range for GRBs detected by \textit{Swift}/BAT is 15 to 150 keV, and that for those detected by \textit{Fermi}/GBM is 10 to 1000 keV.}
\end{TableNotes}
\begin{longtable}{ccccc}
    \caption{Best CHIME/FRB Constraints on Radio Emission from 27 SGRBs\label{table: individual GRBs}}\\
    \midrule \midrule
  \endfirsthead
\bottomrule
\insertTableNotes
\endlastfoot
Name
& Time\tablenotemark{a}
& Flux\tablenotemark{b}
& Fluence Ratio\tablenotemark{c}
& $\eta$\tablenotemark{d}\\
& (hr) & (Jy) & ($10^8$ Jy ms & ($10^{-11}$) \\
& & & erg$^{-1}$ cm$^2$) &
\\ 
\midrule 
GRB 181123B & $10.55$ & $<$1 & $<$0.1 & $<$5\\ 
GRB 181125A & $11.87$ & $<$7000 & $<$200 & $<$8000\\ 
GRB 190326A & $10.92$ & $<$1 & $<$0.1 & $<$4\\ 
GRB 190515A & $-2.88$ & $<$3 & $<$0.3 & $<$10\\ 
GRB 190610A & $6.33$ & $<$20 & $<$0.4 & $<$14\\ 
GRB 190627A & $-5.32$ & $<$20 & $<$3 & $<$120\\ 
GRB 191031D & $2.8$ & $<$3 & $<$0.03 & $<$1.0\\ 
GRB 191106A & $8.63$ & $<$4 & $<$0.2 & $<$8\\ 
GRB 200623A & $2.68$ & $<$7 & $<$0.5 & $<$20\\ 
GRB 200826A & $5.6$ & $<$3 & $<$0.003 & $<$0.1\\ 
GRB 200907B & $-4.07$ & $<$8 & $<$0.6 & $<$20\\ 
GRB 201008A & $6.93$ & $<$3 & $<$0.05 & $<$2\\ 
GRB 201016A & $-5.87$ & $<$7000 & $<$0.7 & $<$30\\ 
GRB 201214B & $-1.22$ & $<$3 & $<$0.1 & $<$5\\ 
GRB 210217A & $5.18$ & $<$2 & $<$0.1 & $<$4\\ 
GRB 210323A & $-4.95$ & $<$2 & $<$0.008 & $<$0.3\\ 
GRB 210413B & $6.9$ & $<$2 & $<$0.09 & $<$4\\ 
GRB 210618A & $4.17$ & $<$2 & $<$0.1 & $<$5\\ 
GRB 210909A & $-2.17$ & $<$5 & $<$0.004 & $<$0.2\\ 
GRB 211023B & $-3.92$ & $<$1 & $<$9 & $<$400\\ 
GRB 211024A & $6.1$ & $<$20 & $<$1 & $<$60\\ 
GRB 220412B & $-1.15$ & $<$6 & $<$0.9 & $<$30\\ 
GRB 220617A & $6.17$ & $<$6 & $<$0.05 & $<$2\\ 
GRB 230205A & $1.93$ & $<$0.5 & $<$0.6 & $<$20\\ 
GRB 230228A & $-5.9$ & $<$400 & $<$1 & $<$50\\ 
GRB 230308A & $11.95$ & $<$600 & $<$0.4 & $<$20\\ 
GRB 230430A & $-1.6$ & $<$3 & $<$0.008 & $<$0.3\\ 
\end{longtable}   
\end{ThreePartTable}
\end{table*}

\subsubsection{Constraints on Radio Luminosities}

In addition to constraining the radio flux for our sample of SGRBs, we can also constrain the radio luminosities for SGRBs with known redshifts. Where possible, the redshifts are taken from the from the Broad-band Repository for Investigating Gamma-ray burst Host galaxies Traits (BRIGHT) catalog \citep{Fong2022, Nugent2022}. For sources that do not have measured redshifts in BRIGHT, we use the redshifts published in GRBWeb. In total, 8 of the 27 SGRBs have published redshifts available either from BRIGHT or GRBWeb. 

\begin{figure*}
  \centering
  \includegraphics[width=0.8\textwidth]{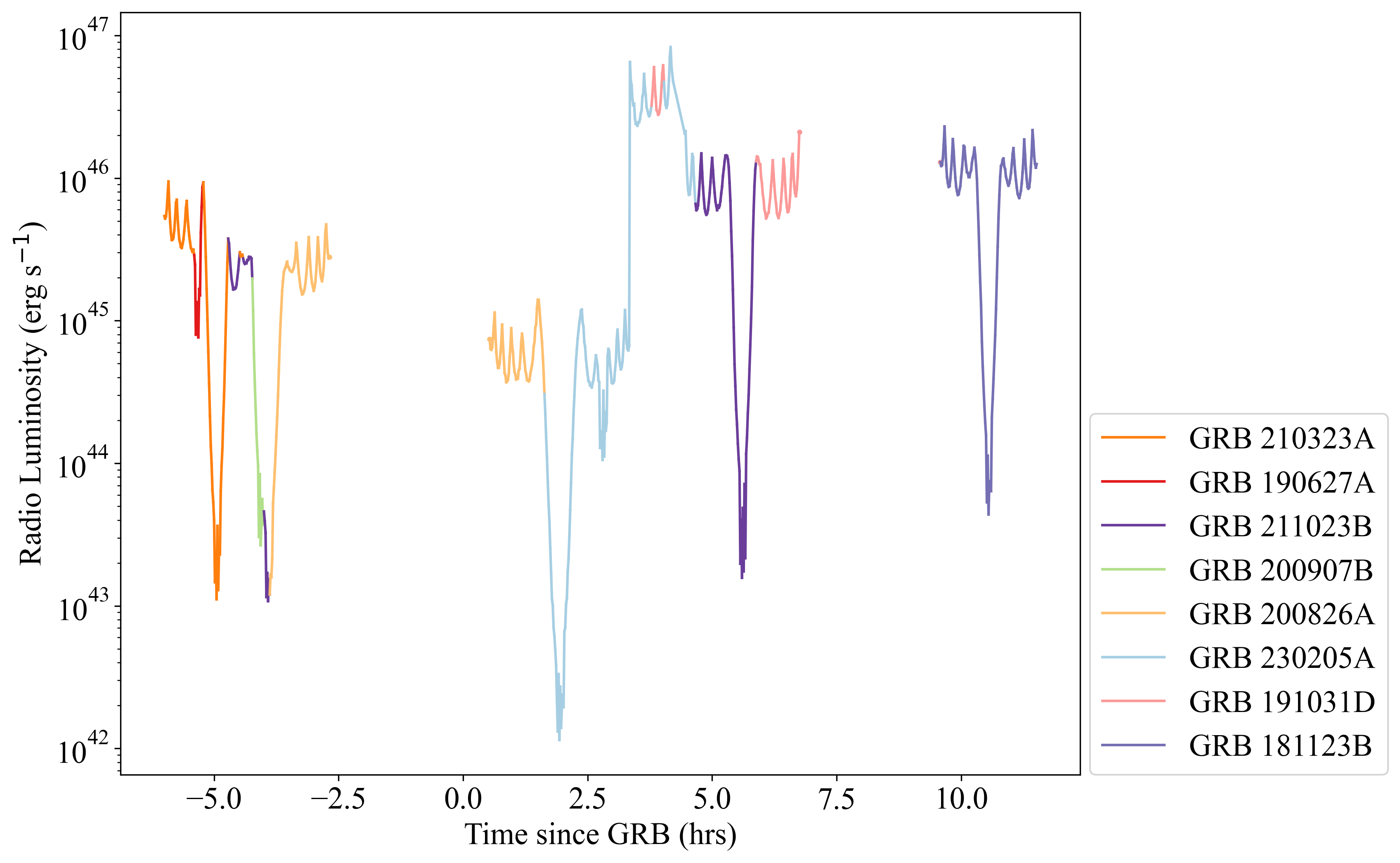}
  \caption{Constraints on the radio luminosity at the 99\% confidence level using the SGRBs within our sample with measured redshifts. Upper limits are calculated every minute starting six hrs prior up until 12 hrs after the SGRB, with the time shifted to account for the estimated DM delay. For each timestamp, we show our most constraining limit from all SGRBs with redshifts that were within the FOV of CHIME/FRB at that relative time. The different SGRBs, along with the time periods over which they are used to constrain the phase space, are shown as different colors.}
  \label{fig:luminosity}
\end{figure*}

In Figure \ref{fig:luminosity}, we show our constraints on the radio luminosities using the 7 SGRBs with redshifts. Luminosities are calculated assuming a flat spectrum with a 400-MHz observing bandwidth e.g., 

\begin{equation}
    L = \frac{\textrm{Flux} \times 4 \pi d_L^2 \times 400 \textrm{MHz}}{(1+z)} \textrm{erg s}^{-1}
\label{Eq: luminosity}
\end{equation}

\noindent
where the flux is in Jy and we use cosmological parameters from \citet{Planck2018} to calculate the luminosity distance, $d_L$. Similar to Figures \ref{fig:fermilimits} and  \ref{fig:swiftlimits}, Figure \ref{fig:luminosity} combines the limits from multiple SGRBs. However, unlike our previous plots, here we combine the limits from \textit{Swift}/BAT and \textit{Fermi}/GBM. 

Unfortunately, neither of the SGRBs for which we can constrain the radio emission at the time of the high-energy emission have measured redshifts. However, we can make assumptions for the redshift of these sources in order to determine a radio luminosity limit at the time of the high-energy emission. To do so, we first construct the SGRB energy distribution using all SGRBs published on GRBWeb that have available redshifts and high-energy fluences. In total, there are 44 SGRBs that meet these criteria. Each SGRB's energy is then calculated using Eq. \ref{Eq: luminosity} by replacing the radio flux limit with the high-energy fluence of the burst. For our constructed energy distribution, the 90th percentile of energies is 
[E$_{\textrm{90th, low}}$, 
E$_{\textrm{90th, high}}$] = 
[$1.2 \times 10^{48}$ erg, 
$6.0 \times 10^{51}$ erg] 
with a mean energy of E$_{\textrm{mean}} = 1.8 \times 10^{50}$ erg. Then, for each of the SGRBs, we use the burst's high-energy fluence to determine a per-source redshift range that corresponds to the above energy values. For example, for GRB 210909A, the high-energy fluence is $3.0 \times 10^{-6}$ erg cm$^{-2}$ and thus the redshifts that correspond to the above energies are 
[z$_{\textrm{low}}$, 
z$_{\textrm{mean}}$, 
z$_{\textrm{high}}] = 
[0.013, 0.16, 0.85$]. 
Finally, for each GRB, we calculate a luminosity constraint assuming z$_{\textrm{low}}$, z$_{\textrm{mean}}$, and z$_{\textrm{high}}$. We list these luminosity constraints, along with the high-energy fluences and inferred redshifts per GRB, in Table \ref{table: Limits At Time GRB}. 

\begin{figure*}
  \centering
  \includegraphics[width=0.9\textwidth]{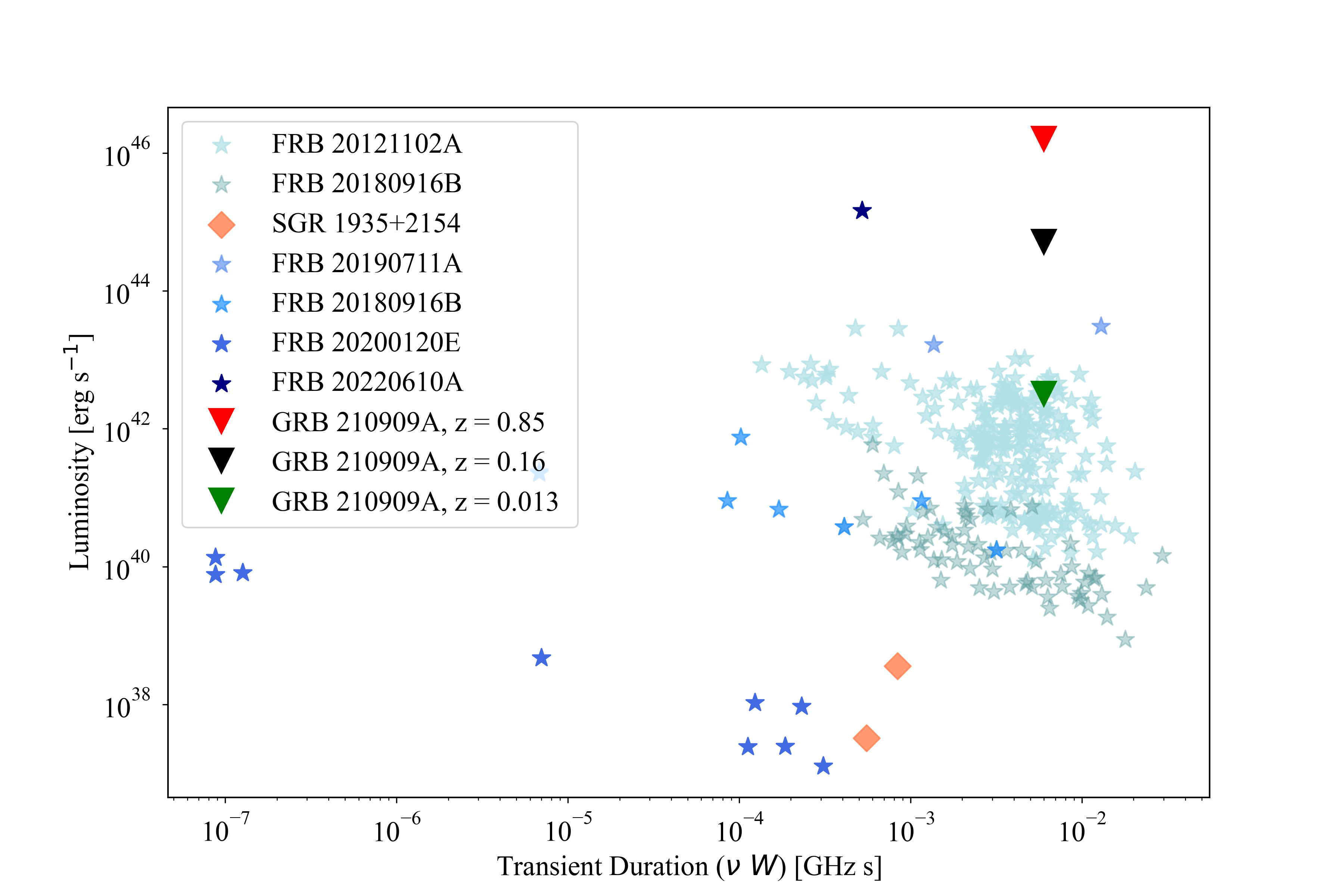}
  \caption{Luminosity vs. frequency-weighted transient duration for multiple radio emitting sources. 
  Bursts from SGR 1935+2154 are shown in purple \citep{2020CHIMESGR, 2020NatureBochenek} and from FRBs in various shades of blue \citep{ssh+16a,sbh+17,lab+17, 2017MNRAS.472.2800H, msh+18, gsp+18, hss+19,hst+19, gms+19, jcf+19, abb+19c, 2020ApJ...897L...4M, rms+20, csa+20, mpm+20, cab+20, mnh+20, aab+20, 2021MNRAS.500.2525K, 2021NatAs...5..594N, 2021ApJ...911L...3P, 2022NatAs...6..393N}. Our most constraining limit at the time of the high-energy emission for GRB 210909A for a redshift of $z = 0.85$ is shown as a red downward arrow, for a redshift of $z = 0.16$ as a black downward arrow, and for a redshift of $z = 0.013$ as a green downward arrow. Given how which we calculate the radio luminosity limits for GRB 210909A, the radio luminosity limits do not depend on the assumed width of the burst. We show the limits for a width of 10 ms, assuming a bandwidth of 400 MHz centered at 600 MHz. However, our luminosity limit applies for multiple burst widths as long as a bandwidth of 400 MHz is assumed. Figure adapted from \citet{2022NatAs...6..393N}.}
  \label{fig:luminosityPhaseSpace}
\end{figure*}

 Our best constraint on the radio luminosity at the time of the high-energy emission is for GRB 210909A for which our flux constraint of $<2000$ Jy corresponds to luminosity constraints of 
 $<3\times10^{42}$ for $z=0.013$, 
 $<5\times10^{44}$ for $z=0.16$, and 
 $<2\times10^{46}$ erg s$^{-1}$ for $z=0.85$.
 In Figure \ref{fig:luminosityPhaseSpace}, we compare the luminosity constraints at the time of the high-energy emission from GRB 210909A with that of bursts from the magnetar SGR 1935+2154 and six different FRBs. Our constraint for a redshift of $z=0.014$ for GRB 210909A is comparable to the brightest bursts from FRB 20181916B and similar to the two bursts from FRB 20190711A \citep{mnh+20, 2021MNRAS.500.2525K}. Our constraint assuming a redshift of $z=0.16$ is $\sim$3 orders of magnitude greater than the average FRB luminosity. \citet{2023Ryder} recently localized FRB 20220610A to a galaxy at a redshift of 1.016, implying a luminosity of $1.6 \times 10^{45}$ erg s$^{-1}$ for this FRB. Given our best flux constraint for GRB 210909A, we can rule out an FRB with a luminosity approx. equal to that of FRB 20220610A out to a redshift of $z\approx0.3$.

\section{Discussion} \label{sec: discussion}

There are multiple models that predict FRB-like radio emission prior to, at the time of, and after the high-energy emission from SGRBs. While \cite{Curtin2023} focus on models that predict radio emission after the high-energy emission, here we focus on models that predict radio emission prior to and at the time of the high-energy emission as this is more in line with predictions from SGRB models. One important caveat to point out prior to discussing the various models is that our constraints only apply if the radio emission is beamed in the same direction as the SGRB. This is a major caveat which we discuss in more detail in Section \ref{subsect: constraint summary}.

\begin{deluxetable*}{l l l l l l l}
\tablecaption{Constraints on Models Predicting Radio Luminosity from Compact Object Mergers} \label{table:model constraints}
\startdata
\\ 
Merger & Reference & $\Delta$t$_{\textrm{radio}}$\tablenotemark{a} & Constraint\tablenotemark{b} & Constraint\tablenotemark{c} & Constraint\tablenotemark{d} & Section\tablenotemark{e} \\
Type & For Model & & ($z=0.013$) & ($z=0.16$) & ($z=0.85$) & \\
\hline
NS-NS & \citet{Sridhar2021Wind} & \begin{tabular}[c]{@{}l@{}}0.02 s prior \\ to 0.1 s post\end{tabular}  & $B_{\textrm{NS}} < 9 \times 10^{13}$ G & $B_{\textrm{NS}} < 2 \times 10^{15}$ G & $B_{\textrm{NS}} < 1 \times 10^{16}$ G& \ref{subsubsect: sridhar} \\
\hline
NS-NS & \citet{Zhang2020} & \begin{tabular}[c]{@{}l@{}}starts $\sim$\\  centuries prior\end{tabular} & $\epsilon < 10^{-12}$\tablenotemark{*} & --- & --- & \ref{subsubsect: zhang 2020}\\
\hline
NS-NS & \citet{Wang2016} & 80 to 2 ms prior & $B_{\textrm{NS}} < 3 \times 10^{12}$ G & $B_{\textrm{NS}} < 4 \times 10^{13}$ G & $B_{\textrm{NS}} < 3 \times 10^{14}$ G & \ref{subsubsect: wang2016} \\
\hline
BH-BH & \citet{Zhang2016BHBH} & $\sim$ ms prior & \begin{tabular}[c]{@{}l@{}} $q < 1 \times 10^{-7}$; $a=1$ \\ $q < 8 \times 10^{-10}$; $a=0.5$ \end{tabular} & \begin{tabular}[c]{@{}l@{}} $q <2\times 10^{-6}$; $a=1$\\  $q < 1 \times 10^{-8}$ ; $a=0.5$ \end{tabular} & \begin{tabular}[c]{@{}l@{}} $q < 1 \times 10^{-5}$; $a=1$\\ $q < 6 \times 10^{-8}$; $a=0.5$ \end{tabular} & \ref{subsubsect: zhang2016} \\
\enddata
\tablenotetext{a}{Predicted time of radio emission relative to the merger.}
\tablenotetext{b}{Parameters from the given model that we constrain using the radio flux limit from GRB 210909A and assuming a redshift of $z=0.013$ for this source. These parameters are described in detail in Section \ref{sec: discussion}. }
\tablenotetext{c}{Same as $b$ except for a redshift of $z=0.16$.}
\tablenotetext{d}{Same as $b$ except for a redshift of $z=0.85$.}
\tablenotetext{e}{Relevant section of the text for discussion of this model and our constraints.}
\tablenotetext{*}{This constraint does not depend on GRB 210909A but instead on GRB 211023B which has a known redshift.}
\end{deluxetable*}

\subsection{NS-NS Mergers}
As previously discussed, SGRBs are produced after the merger of two compact objects. In this section, we focus on SGRBs associated with NS-NS mergers. There are many models that predict radio emission prior to NS-NS binaries but for which the emission would not be observable or detectable by CHIME/FRB. For example, \citet{MostPhilippov2020, 2022MostPhilippov, 2023MostPhilippov} predict radio emission prior to the merger of the NSs due to electromagnetic flares interacting with the orbital current sheet of the binary. However, the FRB-like radio emission would be observed at frequencies much higher than that of CHIME/FRB e.g., $\sim$10 to 20 GHz. In the model by \citet{2020CarrascoShibata}, radio emission could be produced in regions of magnetic reconnection within a current sheet due to collisions of plasmoids. However, the radio flux from this model is estimated to be below the sensitivity limit of CHIME. \citet{HansenLyutikov2001} and \citet{Lyutikov2013} similarly predict radio emission from extraction of the orbital energy of the system through magnetospheric interactions, but again the emission is significantly fainter than our constraints (e.g., L$_{\textrm{radio}}\sim 3 \times 10^{39}$ erg s$^{-1}$). There are, however, a number of models that predict FRB-like radio emission that might be observable by CHIME/FRB. Below, we compare our radio constraints shown in Figures \ref{fig:fermilimits} - \ref{fig:luminosity} to a number of these models. These models, and our constraints, are summarized in Table {\ref{table:model constraints}. 

\subsubsection{FRBs from binary induced winds} \label{subsubsect: sridhar}
\citet{Sridhar2021Wind} predict FRB-like radio emission from an accelerating binary-induced wind produced during the final merger stages of a NS-NS system. In this scenario, one NS is assumed to have a significantly stronger magnetic field than its companion NS \citep[e.g., $\sim 10^{12}$ G for the higher B-field NS and $\sim 10^{10}$ G for the weaker B-field NS;][]{Sridhar2021Wind}. Due to the orbital motion of the system, the (normally) closed magnetic field lines of the more highly magnetized NS would open to infinity around the weaker B-field NS \citep[e.g., see Figure 1 of ][]{Sridhar2021Wind}. This creates an equatorial flow in the orbital plane along these open field lines. If the outflow is accelerating, shocks in the orbital plane could then produce FRBs through the synchrotron maser emission process \citep{mms19}. 

The FRBs would start $\sim$ms prior to the merger\footnote{It is possible that the FRB emission could start earlier if significant shocks are produced during early inspiral stages.} with a peak luminosity at $\sim$0.05-0.1 s post-merger. The radio luminosity would then decay as a function of time. The predicted isotropic radio energy for a coasting fast shell \citep[see][ for details on the definition of a coasting fast shell]{Sridhar2021Wind} is:

\begin{multline}
    E = 1.2 \times 10^{37} \textrm{erg} \left( \frac{\frac{f_{\xi}}{10^{-3}}}{\frac{f_{b}}{0.1}}\right) \left(\frac{\Gamma_f}{10^3}\right)^{4/9} \left(\frac{\nu_{\textrm{obs}}}{1 \textrm{GHz}}\right)^{2/9} \\ \times \left(\frac{B_\textrm{NS}}{10^{12} \textrm{G}}\right)^{16/9}
\label{eq: sridhar}
\end{multline}

\noindent where $f_{\xi}$ is the maser radiative efficiency, $f_{b}$ is the geometric beaming factor, $\Gamma_f$ is the bulk Lorentz factor, $\nu_{\textrm{obs}}$ is the observing frequency, and $B_\textrm{NS}$ is the dipole magnetic field strength of the more magnetized NS \citep[Eq. 35 of ][]{Sridhar2021Wind}. Our best luminosity constraints at the time of a SGRB come from GRB 210909A and are presented in Table \ref{table: Limits At Time GRB}. Assuming a burst width of 10 ms, these luminosity constraints for GRB 210909A correspond to energy constraints of 
$<2 \times 10^{44}$ erg, 
$<5 \times 10^{42}$ erg, and 
$<3 \times 10^{40}$ erg 
assuming redshifts of $z=0.85$, $z=0.16$, and $z=0.013$, respectively, for GRB 210909A. Using Eq. \ref{eq: sridhar} and holding all parameters except for the magnetic field constant, our energy constraints for GRB 210909A imply a magnetic field of 
$B_{\textrm{NS}} < 1 \times 10^{16}$ G, 
$B_{\textrm{NS}} < 2 \times 10^{15}$ G, or $B_{\textrm{NS}} < 9 \times 10^{13}$ G 
assuming redshifts of $z=0.85$, $z=0.16$, or $z=0.013$, respectively. For the high redshift case, our constraint is not particularly interesting. However, for the low redshift case, our B-field constraint is below that of most known magnetars \citep{McGillMagnetarCatalog}.

One major caveat of this model is that the FRB and SGRB arise from different mechanisms, and hence the two transients would not necessarily be beamed in the same direction. The binary wind (and hence the FRB) would likely be beamed along the binary orbital plane, while the GRB jet is typically beamed closer to the binary rotational axis \citep{Sridhar2021Wind, 2014Berger}. Thus, our constraints only apply if the two happen to beamed in the same direction. Radio and optical follow-up of FRB positions to try to identify a radio afterglow or an optical kilonova would instead be better avenues for constraining this model.

\subsubsection{FRBs powered by spin-down energy}\label{subsubsect: zhang 2020}

\citet{Zhang2020} predicts repeating FRBs from the interactions between two NSs' magnetospheres starting $\sim$centuries prior to the merger. In this model, abrupt magnetic reconnections would lead to particle acceleration and hence coherent curvature radiation that would produce FRBs. Using the double-pulsar system PSR J0737$-$3039A/B as a model, \citet{Zhang2020} predicts that the orbital gravitational energy that could be harnessed in a system like this is $\sim 10^{53}$ erg assuming the masses are $1.4$M$_\odot$ with radii of $10^{6}$ cm. While the majority of this would be dissipated in the form of gravitational waves (GWs), it is possible that some might be dissipated in the form of radio emission. 

\citet{Zhang2020} also argues that additional braking could harness the full spin energies of the two NSs which is of order $\sim 10^{45}$ - $10^{49}$ erg.  However, this model does not make any predictions for the radio luminosity as a function of inspiral time, and is rooted largely in observational characteristics of the PSR J0737$-$3039A/B system. Still, given our best pre-SGRB luminosity constraint from GRB 211023B of $\sim10^{43}$ erg s$^{-1}$ at a time $\sim4$ hrs prior to merger (see Figure \ref{fig:luminosity}), we constrain the radio efficiency $\epsilon$ of the orbital gravitational energy to be $<10^{-12}$ if the orbital gravitational energy budget is assumed to be $\sim 10^{53}$ erg and we assume a burst width of 10 ms for GRB 211023B. Additionally, while our energy constraint for GRB 211023B ($<10^{41}$ erg) is below that of the full spin energies, it is likely that not all of the spin energy is harnessed for the FRBs and hence we cannot constrain this aspect of the model.

\subsubsection{FRBs from a unipolar inductor model}\label{subsubsect: wang2016}

\citet{2012Piro}, \citet{Wang2016}, and \citet{Cooper2023} predict that radio emission from a NS-NS merger could be produced through a unipolar inductor model. In this model, one of the NSs has a magnetic field of B $\sim10^{12}$ G while the companion has a significantly weaker (factor of $<100$ weaker) magnetic field. An electromotive force (EMF) is produced when the weaker B-field NS moves through the magnetic field of the higher B-field NS. This EMF then accelerates electrons to ultra-relativistic speeds with FRB-like bursts produced through coherent curvature radiation from these accelerating electrons.

The predicted radio luminosity from \citet{Wang2016} is:

\begin{multline}
    L = 3.5 \times 10^{41} \textrm{erg s}^{-1} \left(\frac{\epsilon}{0.1}\right)^2\left(\frac{B_{\textrm{NS}}}{10^{12} \textrm{G}}\right)^2 \left(\frac{\rho}{30 \textrm{km}}\right)^{-2/3} \\ \times \left(\frac{a}{30 \textrm{km}}\right)^{-8} \left(\frac{\nu}{10^9 \textrm{Hz}}\right)^{1/3}
\label{eq: wang}
\end{multline}

\noindent where $\epsilon$ is the radio efficiency, $B_{\textrm{NS}}$ is the magnetic field of the more strongly magnetized source, $\rho$ is the curvature radius, $a$ is the orbital separation, and $\nu$ is the emitted coherent curvature radiation frequency \citep[See Eq. 16 of ][]{Wang2016}. For a radial separation less than $\sim 30$ km, the radio emission would likely not be able to escape due to absorption, while for a separation greater than $\sim$ 60 km, the radio emission would be too weak to be detectable. 

A full analytical solution to determine the time to coalescence from a separation of $\sim$ 30 km for two NSs is outside the scope of this work. However, we use the estimate from \citet{MaggioreBook} that two 1.4 solar mass objects at a separation of $\sim$33 km will merge within a few ms. Additionally, using Eq. (4.26) from \citet{MaggioreBook}, we scale this to estimate that two objects of 1.4 solar masses at a separation of $\sim$60 km will merge within $\sim$80 ms. Hence, we compare Eq. \ref{eq: wang} with our best limit from the second prior to the merger. Using Eq. \ref{eq: wang} and our best luminosity limit in the $\sim$ second prior to merger, we constrain the surface magnetic field strength to be 
$B_{\textrm{NS}} < 3 \times 10^{14}$ G for a redshift of $z=0.85$,
$B_{\textrm{NS}} < 4 \times 10^{13}$ G for a redshift of $z=0.16$, or
$B_{\textrm{NS}} < 3 \times 10^{12}$ G for a redshift of $z=0.013$. 
While our high redshift constraint is similar to the B-field of most known magnetars, our B-field constraints for $z=0.16$ and $z=0.013$ are below those of most known magnetars \citep{McGillMagnetarCatalog}. This suggests that if  GRB 210909A is a relatively close GRB ($z<0.16$) for which this model applies, and if the radio emission is beamed along our LOS, then the higher B-field NS is likely a regular rotation-powered radio pulsar or a lower B-field NS such as a recycled millisecond pulsar. 

The question of whether the radio emission in the unipolar inductor model would be beamed along our LOS is explored in detail by \citet{Cooper2023}. They find that the radio emission peaks $\sim$10$\degree$ from the magnetic field axis of the primary NS, with emission detectable from $\sim 5-45\degree$ relative to the magnetic axis. From a study of 29 SGRBs, \citet{escorial2022jet} found that the median jet opening angle was 6.1$\degree$, with one opening angle as large as 26$\degree$. Assuming we detect SGRBs on-axis, and that they occur within the main jet opening angle, then a significant portion of the viewable radio emission region could overlap with the viewable SGRB region. However, \citet{Cooper2023} point out that the magnetic field axis of the primary NS is not required to be perpendicular to the orbital plane. Thus, for these configurations, the radio emission would only be observable at large opening angles.

If we were to hold the magnetic field of Eq. \ref{eq: wang} constant at $10^{12}$ G , we could theoretically instead constrain the radio efficiency parameter, $\epsilon$. Assuming a redshift of $z=0.013$, we can only constrain $\epsilon<0.3$. However, $\epsilon$ is expected to be of order $10^{-1}$ or less based on observations of pulsar efficiencies \citep{1993ApJS...88..529T, HansenLyutikov2001, Wang2016}, so this result is not particularly interesting.

One caveat of this model is whether the radio emission would be observable within the band of 400- 800-MHz. \citet{Wang2016} estimate that a photon would need a frequency $\gtrapprox$1 GHz to escape from the surrounding medium. For an emitting frequency of 1GHz, only GRBs with redshifts of $z>0.25$ would have v$_{\textrm{obs}}<800$ MHz. 

\subsection{NS-BH Merger} \label{subsection: wada}
In addition to NS-NS mergers, NS-BH mergers might also produce FRB-like radio emission. In the scenario described by \citet{2023MostPhilippovBHNS}, FRBs would be produced in the $\sim$final orbit of the NS and BH due to the formation of a common magnetosphere. However, similar to their predictions for a NS-NS merger, the radio emission would peak at frequencies much higher than that seen by CHIME/FRB e.g., emitted at 9 GHz and thus even at a redshift of $z=0.85$ observable only at $\sim 4.8$ GHz. \citet{Carrasco2021} similarly study a common magnetosphere produced in a NS-BH merger and predict that FRB-like emission $<1$ GHz will be produced in current sheets outside the light cylinder. The estimated radio luminosity is $\sim 10^{38}$ erg s$^{-1}$ assuming a radio efficiency of $10^{-4}$, significantly below our limits. \citet{Wada2020} also predict that FRBs could be produced during a NS-BH or NS-NS inspiral. Similar to the model by \citet{Carrasco2021}, the inspiral would lead to a spiral arm configuration of current sheets. Magnetic reconnection within these current sheets could produce FRB-like emission. However, in the case of both a NS-NS and NS-BH merger, the predicted radio luminosity is $\sim 10^{36}-10^{39}$ erg s$^{-1}$ (dependent on the assumed radio efficiency of the radiation), again significantly below our limits.

\subsection{BH-BH Merger} \label{subsubsect: zhang2016}
The final model we discuss is that by \citet{Zhang2016BHBH} which predicts FRB-like emission from the merger of two BHs. \citet{Zhang2016BHBH} argues that if at least one of the BHs is charged, then a magnetic dipole perpendicular to the orbital plane could form. As the two BHs inspiral, the magnetic flux would rapidly change, inducing an EMF that could accelerate particles along the field lines and hence produce coherent curvature emission. The peak increase in luminosity in this model would occur in the $\sim$ ms prior to the merger.

The estimated Poynting flux wind luminosity is:

\begin{equation}
L = 1.5 \times 10^{48} \textrm{erg s}^{-1} q^2_{-4} a^{-15}
\label{eq: zhang bh-bh}
\end{equation}

\noindent where $q$ is a dimensionless constant such that the charge of the BH is given by $Q = q Q_c$ with $Q_c$ a characteristic charge given by $Q_c = 2 \sqrt{G M} = (1.0 \times 10^{31}) \frac{M}{10 M_{\odot}}$ e.s.u. and $a$ is a dimensionless quantity such that the orbital separation is given by $a(2r_s)$ with r$_s$ the Schwarzchild radius of the charged BH. The Schwarzchild radius is relevant here as the two BHs are assumed to merge at a separation of $\sim$2r$_s$, although extreme Kerr BHs could merge at a separation of $\sim$r$_s$. Eq. \ref{eq: zhang bh-bh} does not account for the radio efficiency of the emission. We assume an efficiency of $\sim 10^{-3}$ for converting the Poynting flux into radio emission. 

Assuming $a=1$, our best constraints for GRB 210909A imply
$q<3 \times 10^{-4}$, 
$q<6 \times 10^{-5}$ or 
$q<4 \times 10^{-6}$ 
for a redshift of $z=0.85$, $z=0.16$, or $z=0.013$, respectively. 
For a system in which $a=0.5$, we constrain 
$q<2 \times 10^{-6}$, 
$q<3 \times 10^{-7}$ or 
$q<2 \times 10^{-8}$, respectively. 
\citet{Zhang2016BHBH} estimates that $q \approx 10^{-5} - 10^{-4}$ is needed to power a SGRB in this scenario. Our constraints for q for $a=0.5$ are all below below that required to produce a SGRB, while our constraints for $a=1$ are within a magnitude of that needed for a SGRB. Hence, while this model seems unlikely for $a=0.5$ for GRB 210909A, we cannot rule it out for $a=1$.

\subsection{Summary of Constraints and Caveats} \label{subsect: constraint summary}

There are many models predicting FRB-like radio emission from NS-NS or NS-BH mergers that we cannot yet constrain either due to the weak nature of the predicted emission, or the predicted emitted frequency range \citep{HansenLyutikov2001, MostPhilippov2020,  2020CarrascoShibata, Wada2020, Carrasco2021, 2022MostPhilippov, 2023MostPhilippov, 2023MostPhilippovBHNS}. For example, our limits cannot constrain simultaneous radio bursts with luminosities\footnote{Our best luminosity constraint depends on the redshift of GRB 210909A.} $\lessapprox 10^{42}-10^{46}$ erg s$^{-1}$, or bursts for which the emitted frequency would be $\sim$10 GHz. For certain models though \citep[e.g.,][]{Sridhar2021Wind, Zhang2020, Wang2016, Zhang2016BHBH}, we can constrain various aspects of the system such as the NS surface magnetic field strength or the charge of the BH. For example, assuming a redshift of $z=0.013$ for GRB 210909A, our constraints on the NS surface magnetic field in the models of \citet{Wang2016} and \citet{Sridhar2021Wind} are an order of magnitude below the field strengths of most magnetars. This suggests that if GRB 210909A is a relatively close GRB, and if these models apply, then the more highly magnetized NS in the merger is not a high B-field NS. However, many NSs have surface magnetic field strengths $<10^{12}$ G, and hence our constraints still cannot rule out many plausible configurations for a NS-NS merger. Additionally, this requires that GRB 210909a is relatively close e.g., $z=0.013$.

One major caveat is that the FRB may be emitted along a different direction than the SGRB, or have a narrower beaming angle e.g., as discussed in the model by \citet{Sridhar2021Wind}. Thus, we can only draw conclusions about SGRBs and FRBs that are emitted in the same direction. While this caveat warrants a larger discussion, it is outside the scope of this work. One possibility for circumventing this particular caveat and associating an FRB with a compact object merger would instead be to associate an FRB with a GW event \citep{LIGOFRBSearch}. This remains difficult though due to the large uncertainty regions of GW detections and the unknown distances of most FRBs. 

One final caveat is that given the plasma frequency in many of the above models can be $\sim$100s MHz to GHz, we could see a divergence from $\nu^{-2}$ dispersion for the radio waves (assuming the observing frequency is still less than this plasma frequency). However, the dense media surrounding the merger would likely only contribute a very small portion of the total DM, and hence would only contribute a small (likely unobservable) deviation from the classic $\nu^{-2}$ dispersion.

\section{Summary and Future Work}
\label{sec: summary}
In this work, we searched for SGRBs that are temporally and spatially coincident with 4306 CHIME/FRB candidates detected between 2018 July 7 and 2023 August 3. For SGRBs with 1$\sigma$ localization uncertainties $<1\degree$, we do not find any temporally (within 1 week) and spatially (within 3$\sigma$) signficant coincidences. For SGRBs with 1$\sigma$ localization uncertainties $<3\degree$, we also do not find any temporally (within 6 hrs prior through to 12 hrs after the SGRB) and spatially (within 3$\sigma$) significant coincidences. We also search for solely spatial coincidences between all GRBs for which the localization uncertainty is $<1\degree$ and the updated positions from 140 of the FRBs from the first CHIME/FRB catalog \citep{chimefrbcatalog, 2023ChimeBasebandCatalog}. While we find 11 solely spatial coincidences, we conclude that none are statistically significant due to the high chance probability of this occurring. 

Given the lack of spatial and temporal coincidence between our GRB and FRB samples, we calculate upper limits on the possible FRB-like radio emission from 27 SGRBs that were above the horizon at CHIME either prior to, during, or after their high-energy emission. Of the 27 SGRBs, 8 have measured redshifts and thus we can constrain not only their radio flux/fluence but also their radio luminosity. Unfortunately, none of the 8 SGRBs with redshifts were within the FOV of CHIME/FRB at the time of their high-energy emission. Nevertheless, for GRB 210909A, we can constrain the luminosity density to be 
$<2 \times 10^{46}$ erg s$^{-1}$ assuming a redshift $z=0.85$ for the source, 
$<5 \times 10^{44}$ erg s$^{-1}$ assuming a redshift of $z=0.16$, or 
$<3 \times 10^{42}$ erg s$^{-1}$ assuming a redshift $z=0.013$. 
We use these luminosity limits to constrain various parameters from models which predict FRB-like radio emission at the time of the SGRB e.g., the magnetic field of the more highly magnetized NS in a NS-NS merger. 

We will continue to search for CHIME/FRB candidates that are spatially and temporally coincident with GRBs. An exciting new avenue for solely spatial coincidences will also arise with the CHIME/Outriggers, an interferometric set of three CHIME-like telescopes located at continental baselines (Lanman et al. \textit{in prep}). The CHIME/Outrigger telescopes are predicted to be able to localize bright CHIME FRBs to $<50$ mas, allowing for host-identification for hundreds of FRBs. With localization uncertainties of order $\sim$50 mas, solely spatial associations may become possible for these FRBs and GRBs that similarly have very small spatial uncertainties. Additionally, with significantly smaller localization regions, it may become possible to associate FRBs with SGRB afterglows or supernova remenants. This is an exciting future avenue for investigating possible progenitors for FRBs. 


\section*{acknowledgments}
We thank N. Sridhar, A. Tohuvavohu, and P. Coppin for helpful comments during the writing of this manuscript.

We acknowledge the use of public data from the \textit{Swift} data archive, the \textit{Fermi} data archive, GRBWeb, and the GCN Notices. We acknowledge that CHIME is located on the traditional, ancestral, and unceded territory of the Syilx/Okanagan people. We are grateful to the staff of the Dominion Radio Astrophysical Observatory, which is operated by the National Research Council of Canada. CHIME is funded by a grant from the Canada Foundation for Innovation (CFI) 2012 Leading Edge Fund (Project 31170) and by contributions from the provinces of British Columbia, Qu\'{e}bec and Ontario. The CHIME/FRB Project is funded by a grant from the CFI 2015 Innovation Fund (Project 33213) and by contributions from the provinces of British Columbia and Qu\'{e}bec, and by the Dunlap Institute for Astronomy and Astrophysics at the University of Toronto. Additional support was provided by the Canadian Institute for Advanced Research (CIFAR), McGill University and the Trottier Space Institute thanks to the Trottier Family Foundation, and the University of British Columbia. 

\allacks

\bibliography{frbrefs, refs}
\bibliographystyle{aasjournal}

\appendix

\section{Upper Limits as a Function of Time for SGRBs and LGRBs}
\label{sec:Appendix on UL tables}

Below, in Tables \ref{table: SGRBs Swift} and \ref{table: SGRBs Fermi}, we present the radio flux and fluence ratio limits that are shown in Figure \ref{fig:fermilimits} and  Figure \ref{fig:swiftlimits}. For each time bin (each 0.25 hr), we present our most constraining radio flux limits and fluence ratios. For each timestamp, we give the range of our most constraining upper limits calculated considering the entire sample of GRBs within that category (e.g., for Table \ref{table: SGRBs Swift}, all SGRBs detected by \textit{Swift}/BAT) are used to determine the limits. Additionally, in Table \ref{table: Luminosity constraints}, we present the luminosity limits shown in Figure \ref{fig:luminosity}.

\begin{ThreePartTable}
\begin{TableNotes}
\item\tablenotetext{a}{End time of the given time bin (in hrs) for which the radio flux/fluence limit ranges apply. Times are relative to the detected high-energy emission, with negative times indicating emission prior to the high-energy emission. Start time for the range of fluxes/fluences is 0.25 hr prior to the end time. These times correct for an estimated dispersion delay (e.g., see Section \ref{subsec: CalcUL}).}

\tablenotetext{b}{Range of upper limits on the possible radio flux at the 99$\%$ confidence level for a 10-ms radio burst for the given time bin. The flux range is calculated using all \textit{SGRBs} detected by \textit{Swift}/BAT within our sample.}

\tablenotetext{c}{Same as $b$ except for the radio-to-high-energy fluence ratio.}

\tablenotetext{d}{Same as $b$ except for $\eta$ (dimensionless radio-to-high-energy fluence ratio assuming a 400-MHz radio emission bandwidth). The \textit{Swift}/BAT high-energy band is 15 to 150 keV, while the CHIME/FRB radio band is 400- to 800-MHz.}
\end{TableNotes}
\begin{longtable}{cccc}
    \caption{CHIME/FRB Upper Limits on Radio Emission from SGRBs detected by  \textit{Swift}/BAT}
    \label{table: SGRBs Swift}\\
    \midrule \midrule
  \endfirsthead

\bottomrule
\insertTableNotes
\endlastfoot

Time \tablenotemark{a}
& Flux \tablenotemark{b}
& Fluence Ratio \tablenotemark{c}
& $\eta$\tablenotemark{d} \\
(hr) & (Jy) & ($10^8$ Jy ms & ($10^{-11}$)\\
 & & erg$^{-1}$ cm$^2)$ &
\\
\midrule
$-5.75$ & $<$3000-7000 & $<$400-1100 & $<$20000-40000\\ 
$-5.5$ & $<$800-8000 & $<$130-1000 & $<$5000-40000\\ 
$-5.25$ & $<$20-500 & $<$3-70 & $<$120-3000\\ 
$-5.0$ & $<$200-2000 & $<$30-300 & $<$1400-13000\\ 
$-4.75$ & $<$200-500 & $<$130-200 & $<$5000-8000\\ 
$-4.5$ & $<$200-500 & $<$130-200 & $<$5000-8000\\ 
$-4.25$ & $<$300-500 & $<$70-300 & $<$3000-13000\\ 
$-4.0$ & $<$6-300 & $<$0.6-40 & $<$20-1700\\ 
$-3.75$ & $<$1-20 & $<$4-130 & $<$170-5000\\ 
$-3.5$ & $<$30-300 & $<$120-200 & $<$5000-10000\\ 
$-3.25$ & $<$200-400 & $<$120-300 & $<$5000-11000\\ 
$-3.0$ & $<$200-500 & $<$150-3000 & $<$6000-130000\\ 
$-2.75$ & $<$200-600 & $<$1500-4000 & $<$60000-160000\\ 
$-2.5$ & $<$300-400 & $<$2000-3000 & $<$90000-120000\\ 
$0.75$ & $<$200-400 & $<$200-600 & $<$9000-20000\\ 
$1.0$ & $<$150-400 & $<$200-500 & $<$7000-20000\\ 
$1.25$ & $<$200-300 & $<$200-400 & $<$8000-20000\\ 
$1.5$ & $<$100-600 & $<$8-200 & $<$300-8000\\ 
$1.75$ & $<$10-500 & $<$7-20 & $<$300-600\\ 
$2.0$ & $<$0.5-7 & $<$0.6-8 & $<$20-300\\ 
$2.25$ & $<$0.8-200 & $<$1.0-20 & $<$40-700\\ 
$2.5$ & $<$200-500 & $<$4-20 & $<$140-700\\ 
$2.75$ & $<$3-100 & $<$0.03-3 & $<$1-100\\ 
$3.0$ & $<$3-60 & $<$0.03-0.6 & $<$1.0-20\\ 
$3.25$ & $<$90-500 & $<$0.8-13 & $<$30-500\\ 
$3.5$ & $<$200-1100 & $<$6-20 & $<$200-600\\ 
$3.75$ & $<$500-1000 & $<$6-13 & $<$200-500\\ 
$4.0$ & $<$500-1000 & $<$7-15 & $<$300-600\\ 
$4.25$ & $<$500-1400 & $<$8-70 & $<$300-3000\\ 
$4.5$ & $<$800-900 & $<$50-60 & $<$2000-2000\\ 
$4.75$ & $<$120-800 & $<$8-50 & $<$300-2000\\ 
$5.0$ & $<$8-100 & $<$0.5-6 & $<$20-200\\ 
$5.25$ & $<$2-7 & $<$0.1-0.4 & $<$4-20\\ 
$5.5$ & $<$3-30 & $<$0.2-2 & $<$7-70\\ 
$5.75$ & $<$30-700 & $<$2-50 & $<$80-2000\\ 
$6.0$ & $<$700-800 & $<$50-50 & $<$2000-2000\\ 
$6.25$ & $<$200-1100 & $<$3-60 & $<$130-2000\\ 
$6.5$ & $<$20-800 & $<$0.4-20 & $<$14-700\\ 
$6.75$ & $<$400-1000 & $<$30-60 & $<$1100-2000\\ 
$7.0$ & $<$500-1000 & $<$30-70 & $<$1100-3000\\ 
$7.25$ & $<$500-1100 & $<$30-70 & $<$1200-3000\\ 
$7.5$ & $<$400-900 & $<$20-50 & $<$800-2000\\ 
$7.75$ & $<$300-800 & $<$20-50 & $<$800-2000\\ 
$8.0$ & $<$300-700 & $<$20-40 & $<$700-2000\\ 
$8.25$ & $<$500-800 & $<$30-40 & $<$1200-2000\\ 
$8.5$ & $<$13-500 & $<$0.8-30 & $<$30-1100\\ 
$8.75$ & $<$4-20 & $<$0.2-1 & $<$8-50\\ 
$9.0$ & $<$30-500 & $<$1-30 & $<$60-1100\\ 
$9.25$ & $<$400-700 & $<$30-40 & $<$1000-2000\\ 
$9.5$ & $<$300-700 & $<$20-40 & $<$700-2000\\ 
$9.75$ & $<$300-800 & $<$20-40 & $<$700-2000\\ 
$10.0$ & $<$200-800 & $<$20-50 & $<$800-2000\\ 
$10.25$ & $<$300-500 & $<$30-40 & $<$1000-1800\\ 
$10.5$ & $<$5-500 & $<$0.4-40 & $<$16-1700\\ 
$10.75$ & $<$1-60 & $<$0.1-4 & $<$5-150\\ 
$11.0$ & $<$1-18 & $<$0.1-1 & $<$4-40\\ 
$11.25$ & $<$13-400 & $<$0.8-40 & $<$30-1400\\ 
$11.5$ & $<$300-600 & $<$20-50 & $<$900-2000\\ 
$11.75$ & $<$400-900 & $<$20-50 & $<$1000-2100\\
\end{longtable}   
\end{ThreePartTable}

\FloatBarrier

\begin{ThreePartTable}
\begin{TableNotes}
\item\tablenotetext{a}{End time of the given time bin (in hrs) for which the radio flux/fluence limit ranges apply. Times are relative to the detected high-energy emission, with negative times indicating emission prior to the high-energy emission. Start time for the range of fluxes/fluences is 0.25 hr prior to the end time. These times correct for an estimated dispersion delay (e.g., see Section \ref{subsec: CalcUL}).}

\tablenotetext{b}{Range of upper limits on the possible radio flux at the 99$\%$ confidence level for a 10-ms radio burst for the given time bin. The flux range is calculated using all SGRBs detected by \textit{Fermi}/GBM within our sample.}

\tablenotetext{c}{Same as $b$ except for the radio-to-high-energy fluence ratio.}

\tablenotetext{d}{Same as $b$ except for $\eta$ (dimensionless radio-to-high-energy fluence ratio assuming a 400-MHz radio emission bandwidth). The \textit{Fermi}/GBM high-energy band is 10 to 1000 keV, while the CHIME/FRB radio band is 400- to 800-MHz.}
\end{TableNotes}

\begin{longtable}{cccc}
    \caption{CHIME/FRB Upper Limits on Radio Emission from SGRBs detected by \textit{Fermi}/GBM}
    \label{table: SGRBs Fermi}\\
    \midrule \midrule
  \endfirsthead

\bottomrule
\insertTableNotes
\endlastfoot

Time \tablenotemark{a}
& Flux \tablenotemark{b}
& Fluence Ratio \tablenotemark{c}
& $\eta$\tablenotemark{d} \\
(hr) & (Jy) & ($10^8$ Jy ms & ($10^{-11}$)\\
 & & erg$^{-1}$ cm$^2)$ &
\\
\midrule
$-5.75$ & $<$400-1200 & $<$0.7-2 & $<$30-80\\ 
$-5.5$ & $<$500-1200 & $<$0.9-6 & $<$40-300\\ 
$-5.25$ & $<$500-2000 & $<$3-5 & $<$110-200\\ 
$-5.0$ & $<$7-2000 & $<$0.03-6 & $<$1-300\\ 
$-4.75$ & $<$2-300 & $<$0.008-1 & $<$0.3-50\\ 
$-4.5$ & $<$400-2000 & $<$2-7 & $<$80-300\\ 
$-4.25$ & $<$500-1100 & $<$2-3 & $<$80-100\\ 
$-4.0$ & $<$500-1300 & $<$1-3 & $<$50-100\\ 
$-3.75$ & $<$500-1300 & $<$0.4-1.0 & $<$14-40\\ 
$-3.5$ & $<$200-500 & $<$0.2-0.5 & $<$6-20\\ 
$-3.25$ & $<$200-300 & $<$0.1-0.3 & $<$5-10\\ 
$-3.0$ & $<$15-200 & $<$0.05-0.1 & $<$2-5\\ 
$-2.75$ & $<$3-30 & $<$0.02-0.05 & $<$0.9-2\\ 
$-2.5$ & $<$8-30 & $<$0.006-0.02 & $<$0.2-0.9\\ 
$-2.25$ & $<$6-30 & $<$0.005-0.02 & $<$0.2-0.8\\ 
$-2.0$ & $<$5-20 & $<$0.004-0.02 & $<$0.2-0.8\\ 
$-1.75$ & $<$6-20 & $<$0.004-0.02 & $<$0.2-0.8\\ 
$-1.5$ & $<$3-30 & $<$0.008-0.03 & $<$0.3-1\\ 
$-1.25$ & $<$13-70 & $<$0.02-0.07 & $<$0.9-3\\ 
$-1.0$ & $<$3-300 & $<$0.07-0.2 & $<$3-9\\ 
$-0.75$ & $<$200-400 & $<$0.1-0.3 & $<$4-12\\ 
$-0.5$ & $<$200-600 & $<$0.2-0.5 & $<$7-20\\ 
$-0.25$ & $<$500-3000 & $<$0.4-2 & $<$16-90\\ 
$0.0$ & $<$2000-2000 & $<$2-2 & $<$60-80\\ 
$0.25$ & $<$2000-7000 & $<$2-6 & $<$80-200\\ 
$0.5$ & $<$5000-9000 & $<$4-8 & $<$200-300\\ 
$0.75$ & $<$5000-11000 & $<$4-9 & $<$200-400\\ 
$1.0$ & $<$4000-11000 & $<$3-9 & $<$100-400\\ 
$1.25$ & $<$1000-4000 & $<$3-8 & $<$100-300\\ 
$1.5$ & $<$800-2000 & $<$3-4 & $<$110-200\\ 
$1.75$ & $<$800-2000 & $<$2-6 & $<$90-200\\ 
$2.0$ & $<$700-2000 & $<$2-5 & $<$90-200\\ 
$2.25$ & $<$1000-2000 & $<$2-4 & $<$80-200\\ 
$2.5$ & $<$50-1200 & $<$2-5 & $<$70-200\\ 
$2.75$ & $<$7-40 & $<$0.5-3 & $<$20-120\\ 
$3.0$ & $<$10-500 & $<$1.0-4 & $<$40-200\\ 
$3.25$ & $<$300-1100 & $<$2-4 & $<$70-200\\ 
$3.5$ & $<$300-700 & $<$2-4 & $<$60-200\\ 
$3.75$ & $<$400-1100 & $<$2-4 & $<$60-200\\ 
$4.0$ & $<$60-700 & $<$2-4 & $<$70-200\\ 
$4.25$ & $<$2-40 & $<$0.1-2 & $<$5-60\\ 
$4.5$ & $<$6-500 & $<$0.3-3 & $<$12-100\\ 
$4.75$ & $<$400-600 & $<$1.0-2 & $<$40-80\\ 
$5.0$ & $<$300-600 & $<$0.9-3 & $<$40-100\\ 
$5.25$ & $<$300-700 & $<$1.0-2 & $<$40-80\\ 
$5.5$ & $<$30-800 & $<$0.03-2 & $<$1-70\\ 
$5.75$ & $<$3-70 & $<$0.003-0.08 & $<$0.1-3\\ 
$6.0$ & $<$110-600 & $<$0.1-2 & $<$5-100\\ 
$6.25$ & $<$6-90 & $<$0.05-1 & $<$2-50\\ 
$6.5$ & $<$30-500 & $<$0.3-2 & $<$10-90\\ 
$6.75$ & $<$7-200 & $<$0.3-3 & $<$10-100\\ 
$7.0$ & $<$2-5 & $<$0.05-0.2 & $<$2-8\\ 
$7.25$ & $<$3-400 & $<$0.06-3 & $<$2-110\\ 
$7.5$ & $<$400-500 & $<$2-3 & $<$80-100\\ 
$7.75$ & $<$300-500 & $<$1.0-3 & $<$40-100\\ 
$8.0$ & $<$300-500 & $<$0.3-0.9 & $<$14-40\\ 
$8.25$ & $<$400-500 & $<$0.3-0.4 & $<$12-20\\ 
$8.5$ & $<$110-500 & $<$0.08-0.5 & $<$3-20\\ 
$8.75$ & $<$40-120 & $<$0.03-0.1 & $<$1-4\\ 
$9.0$ & $<$30-40 & $<$0.02-0.03 & $<$0.8-1\\ 
$9.25$ & $<$30-30 & $<$0.02-0.02 & $<$0.8-1.0\\ 
$9.5$ & $<$8-30 & $<$0.006-0.02 & $<$0.2-0.9\\ 
$9.75$ & $<$9-30 & $<$0.007-0.02 & $<$0.3-1.0\\ 
$10.0$ & $<$9-30 & $<$0.007-0.02 & $<$0.3-1.0\\ 
$10.25$ & $<$8-30 & $<$0.006-0.03 & $<$0.2-1\\ 
$10.5$ & $<$11-30 & $<$0.008-0.03 & $<$0.3-1\\ 
$10.75$ & $<$30-40 & $<$0.02-0.03 & $<$0.9-1\\ 
$11.0$ & $<$40-110 & $<$0.03-0.09 & $<$1-4\\ 
$11.25$ & $<$90-300 & $<$0.07-0.2 & $<$3-9\\ 
$11.5$ & $<$400-700 & $<$0.3-0.5 & $<$11-20\\ 
$11.75$ & $<$600-1000 & $<$0.4-0.7 & $<$20-30\\ 
\end{longtable}   
\end{ThreePartTable}

\begin{ThreePartTable}
\begin{TableNotes}
\item\tablenotetext{a}{End time of the given time bin (in hrs) for which the radio luminosity constraint ranges apply. Times are relative to the detected high-energy emission, with negative times indicating emission prior to the high-energy emission. Start time for the range of fluxes/fluences is 0.25 hr prior to the end time.}

\tablenotetext{b}{Range of upper limits on the possible radio luminosity at the 99$\%$ confidence level for a 10-ms radio burst for the given time bin. The flux range is calculated using all SGRBs detected by \textit{Fermi}/GBM and \textit{Swift}/BAT within our sample.}
\end{TableNotes}

\begin{longtable}{cc}
    \caption{CHIME/FRB Constraints on SGRB Luminosities}
    \label{table: Luminosity constraints}\\
    \midrule \midrule
  \endfirsthead

\bottomrule
\insertTableNotes
\endlastfoot

Time \tablenotemark{a}
& Luminosity Range \tablenotemark{b}\\
(hr) & ($10^{45}$ erg s$^{-1}$)
\\
\midrule
$-5.75$ & $<$3 - 9\\ 
$-5.5$ & $<$3 - 6\\ 
$-5.25$ & $<$0.8 - 6\\ 
$-5.0$ & $<$0.04 - 9\\ 
$-4.75$ & $<$0.01 - 1\\ 
$-4.5$ & $<$1 - 3\\ 
$-4.25$ & $<$2 - 3\\ 
$-4.0$ & $<$0.03 - 2\\ 
$-3.75$ & $<$0.01 - 0.2\\ 
$-3.5$ & $<$0.2 - 2\\ 
$-3.25$ & $<$1 - 3\\ 
$-3.0$ & $<$1 - 3\\ 
$-2.75$ & $<$1 - 4\\ 
$-2.5$ & $<$2 - 3\\ 
$0.75$ & $<$0.5 - 1\\ 
$1.0$ & $<$0.4 - 0.9\\ 
$1.25$ & $<$0.4 - 0.8\\ 
$1.5$ & $<$0.4 - 1\\ 
$1.75$ & $<$0.03 - 1\\ 
$2.0$ & $<$0.0 - 0.02\\ 
$2.25$ & $<$0.0 - 0.4\\ 
$2.5$ & $<$0.4 - 1\\ 
$2.75$ & $<$0.1 - 0.6\\ 
$3.0$ & $<$0.1 - 0.6\\ 
$3.25$ & $<$0.4 - 1\\ 
$3.5$ & $<$0.6 - 65\\ 
$3.75$ & $<$25 - 54\\ 
$4.0$ & $<$28 - 60\\ 
$4.25$ & $<$31 - 83\\ 
$4.5$ & $<$10 - 21\\ 
$4.75$ & $<$6 - 15\\ 
$5.0$ & $<$5 - 15\\ 
$5.25$ & $<$6 - 13\\ 
$5.5$ & $<$0.2 - 14\\ 
$5.75$ & $<$0.02 - 0.5\\ 
$6.0$ & $<$0.7 - 14\\ 
$6.25$ & $<$5 - 13\\ 
$6.5$ & $<$5 - 14\\ 
$6.75$ & $<$5 - 21\\ 
$9.75$ & $<$8 - 23\\ 
$10.0$ & $<$7 - 19\\ 
$10.25$ & $<$10 - 17\\ 
$10.5$ & $<$0.2 - 16\\ 
$10.75$ & $<$0.04 - 5\\ 
$11.0$ & $<$7 - 14\\ 
$11.25$ & $<$7 - 16\\ 
$11.5$ & $<$8 - 22\\ 
\end{longtable}   
\end{ThreePartTable}

\FloatBarrier

\end{document}